\documentclass{article}
\usepackage{graphicx} 
\usepackage{appendix}
\usepackage{amsmath}
\usepackage{amssymb}
\usepackage{slashed}
\usepackage{hyperref}
\usepackage{braket}
\usepackage{comment}
\usepackage{ulem}
\usepackage{CJK}
\usepackage{dirtytalk}
\usepackage{authblk}
\usepackage[numbers,square,super,sort&compress]{natbib}
\newcommand{\acal}{\mathcal{A}}

\newcommand{\lcal}{\mathcal{L}}

\newcommand{\fcal}{\mathcal{F}}

\newcommand{\ccal}{\mathcal{C}}

\newcommand{\cbb}{\mathbb{C}}
\newcommand{\rbb}{\mathbb{R}}

\newcommand{\gfrak}{\mathfrak{g}}

\title{New edge modes and corner charges for first-order symmetries of 4D gravity}

\author[1]{Simon Langenscheidt}
\author[2]{Daniele Oriti}
\affil[1]{\small Arnold Sommerfeld Center for Theoretical Physics, Ludwig-Maximilians-Universit\"at M\"unchen, Theresienstrasse 37, 80333 M\"unchen, Germany ; Munich Center for Quantum Science and Technology (MCQST), Schellingstrasse 4, 80799 M\"unchen, Germany}
\affil[2]{Depto. de F\'isica Te\'orica, Facultad de Ciencias F\'isicas, Universidad Complutense de Madrid, Plaza de las Ciencias 1, 28040 Madrid, Spain, EU ;
 Munich Center for Quantum Science and Technology (MCQST), Schellingstrasse 4, 80799 M\"unchen, Germany; 
 Department of Physics, Shanghai University, 99 Shangda Rd, 200444, Shanghai, P.R.China}

\date{March 2024}

\begin{document}

\maketitle
\begin{abstract}
    We present a set of noncommuting frame-translation symmetries in 4D gravity in tetrad-connection variables, which allow expressing diffeomorphisms as composite transformations. Working on the phase space level for finite regions, we pay close attention to the corner piece of the generators, discuss various possible charge brackets, relative definitions of the charges, coupling to spinors and relations to other charges. 
    What emerges is a picture of the symmetries and edge modes of gravity that bears local resemblance to a Poincare group $SO(1,3)\ltimes \mathbb{R}^{1,3}$, but possesses structure functions. In particular, we argue that the symmetries and charges presented here are more amenable to discretisation, and sketch a strategy for this charge algebra, geared toward quantum gravity applications.
    
\end{abstract}

\tableofcontents

\section{Introduction}
    In the search for quantum formulations of gravitational physics, several recent approaches emphasize the role of special transformations of the fields known as corner symmetries\cite{Freidel:2015gpa,Chandrasekaran:2021vyu,gomesUnifiedGeometricFramework2019,freidelCornerSymmetryQuantum2023,Ball:2024hqe,carrozzaEdgeModesReference2022,Carrozza:2022xut,freidelEdgeModesGravity2020,freidelEdgeModesGravity2020a,freidelExtendedCornerSymmetry2021,donnellyGravitationalEdgeModes2021,freidelGravitationalEnergyLocal2015,ciambelliIsolatedSurfacesSymmetries2021a,donnellyLocalSubsystemsGauge2016,kabelQuantumReferenceFrames2023}. These transformations, coming in typically two flavors respectively at finite distance and asymptotic boundaries, arise from the intrinsic structure of most gauge theories. In principle, a gauge symmetry, such as diffeomorphisms in gravity, is conventionally labeled a redundancy of the theory. This is due to the fact that the free time-dependence of transformation parameters can change any solutions of the (classical) theory into another solution with the \textit{same} initial data; this makes the time evolution indeterministic at face value.\\
    In this sense, it appears that one can readily discard the physical content of gauge transformations. However, when the Cauchy slice $\Sigma$ that is being evolved has a boundary (or corner) $\partial\Sigma$ (asymptotic or finite distance), gauge transformations with support on it are \textit{physical transformations} that permute between states. These corner symmetries are at the heart of why gauge theories featuring diffeomorphism invariance (such as gravity, BF theory or Chern-Simons theory) all have vastly larger state spaces when boundaries are present. \\
    From this perspective alone, it is worth to closely investigate the full set of corner symmetries admitted by gravity in aid of finding interesting sets of observables, which, in the pure bulk theory, are rare at best. 
     The asymptotic case of these symmetries has seen intense research activity both in the Anti-deSitter\cite{Brown:1986nw,coussaertAsymptoticDynamicsThreedimensional1995,Compere:2008us,Compere:2020lrt} and asymptotically flat\cite{donnayBMSFluxAlgebra2021,barnichSymmetriesAsymptoticallyFlat2010,freidelWeylBMSGroup2021a} sectors of General Relativity (GR), where asymptotic diffeomorphism symmetries allow for the organisation, analysis and even reconstruction of several elements of gravitational physics.
    On the other hand, in the finite distance case, particularly in 3D gravity, research has focused on the use of finite distance symmetry charges to aid quantisation\cite{freidel1DLoopQuantum2019,geiller3dGravityBondiWeyl2021a,geillerDualDiffeomorphismsFinite2022a,freidelQuantumGravityDisk2021}. \\
    The main idea of this approach, sometimes referred to as local holography, is the reconstruction of small bulk regions from corner data alone in a similar spirit to the AdS/CFT correspondence, but at finite distance. \\
    Upon achieving such (re-)constructions, larger regions may be built in several ways, the most conservative being a direct glueing of spatial or spatiotemporal regions with appropriate fusion rules. Alternatively, one might forego the bulk description for a picture of corner degrees of freedom as 'molecules', whose hydrodynamic, macroscopic dynamics is possibly equivalent to the continuum bulk physics.\\
    Additionally, even without the microscopic holography perspective, one nowadays appreciates the fact that any conventional calculations of von Neumann entropies in gauge theories include contributions from edge mode degrees of freedom\cite{Freidel:2015gpa}.\\
    Therefore, it is crucial to understand precisely which gauge symmetries, and therefore edge modes, gravity has, and establish a link between the work done in the 3D and the 4D case. Here, we do so by providing a 4D analogue of an internal gauge symmetry of 3D gravity that can be understood as being a constituent of diffeomorphisms.
    Let us illustrate these symmetries here.\\
    In the tetrad formulation in 3D\cite{geillerMostGeneralTheory2021a}, we begin from the action (see appendix \ref{App:Conventions} for notation)
    \begin{equation}
        S_{3D} = \int_M (\star\theta)\wedge F_\omega
    \end{equation}
    modeled by a triad $\theta$, a spin connection $\omega$ for $SO(1,2)$ and $\star:\rbb^{1,2}\rightarrow so(1,2)$ denoting the internal dual. This action, being of BF-type, enjoys 2 sets of symmetries known as Lorentz transformations
    \begin{equation}
        X_\alpha[\omega] = d_\omega\alpha \qquad X_\alpha[\theta]=-\alpha\cdot\theta \quad \alpha\in \Omega^{0}(M,\gfrak),
    \end{equation}
    and Kalb-Ramond translations
    \begin{equation}
        Y_\mu[\omega] = 0\qquad Y_\mu[\theta]= d_\omega\mu\quad \mu \in\Omega^{1}(M,\gfrak).
    \end{equation}
    On the phase space of a slice $\Sigma$, one can associate generators to these symmetries which encapsulate a lot of the degrees of freedom of finite regions in 3D gravity; respectively,
    \begin{equation}
        \begin{gathered}
            C_\alpha = \int_\Sigma \star\theta\wedge d_\omega\alpha \approx \oint_{\partial\Sigma} \star\theta\cdot\alpha\\
            K_\mu = -\int_\Sigma \mu\cdot F_\omega +
            \oint_{\partial\Sigma}
            \mu\cdot\omega
            \approx \oint_{\partial\Sigma}
            \mu\cdot\omega
        \end{gathered}
    \end{equation}
    where weak equality $\approx$ indicates using the constraints of the theories, given respectively by
    \begin{equation}
        d_\omega \theta\approx 0\approx F_\omega.
    \end{equation}
    Given this set of transformations, diffeomorphisms of 3D gravity can be understood as effective combinations of these two, and charges thereof are quadratic combinations of these more fundamental symmetries.\cite{geillerDiffeomorphismsQuadraticCharges2022}\\
    This offers a useful way to bring diffeomorphism invariance also to the discrete level, as exemplified in state sum models for 3D gravity such as the Ponzano-Regge model.\cite{baratinDiffeomorphismsGroupField2011}
 In these models, diffeomorphism invariance is commonly understood through triangulation invariance, which in turn is captured by invariance under the 3D Pachner moves. Furthermore, it has been demonstrated that this invariance can be understood as a \textit{geometric} one where vertices of a given triangulation may be freely translated. This, however, can be expressed through labels of the  triangulation only, and therefore makes no reference to a background embedding of the triangulation cells. In this way, one can \textit{abstractly} realise the action of diffeomorphisms on discretised gravity in such a way that no mention of points or vector fields generating the diffeomorphisms is necessary. \\
    Therefore, a natural question arises about the existence of similar transformations in 4D gravity in tetrad variables. While Lorentz transformations are perfectly well-known, the case of translation charges is much less obvious. For example, it is well-known\cite{dittrichDiffeomorphismSymmetryQuantum2008} that Regge calculus, as a discrete path integral for the metric, admits a \textit{partial} vertex translation symmetry akin to the one found in 3D, but this pertains only to the flat sector. Quite clearly, the transformation that would replace the 3D symmetry must be modified to include effects of curvature and torsion.\\
    It is the goal of the present note to elaborate on this issue and propose a useful generalisation of the 3-dimensional translation charges. To begin, we review the phase space structure of gravity in tetrad variables, highlighting some important complications that require us to restrict the set of dynamical fields on the phase space. Then, with this background, we present phase space nonlinear Kalb-Ramond translations for 4D gravity, and discuss their complicated relationship with transformations of the spacetime fields. Further, we consider two obvious corner extensions of their generators and discuss the relation to Brown-York charges before evaluating them in the simple case of a Schwarzschild geometry. Finally, we give an extended discussion in which we focus on the effects we can expect on edge modes, together with modifications due to the inclusion of matter. We close with an outlook on several possible applications of the charges.

\section{The phase space of gravity in tetrad variables}
    In this section, we provide necessary technical setup: We specify the set of fields, its canonical structure and highlight that the symplectic form is degenerate, leading to the crucial requirement of fixing the so-called 'structural gauge'. This is in stark contrast to the case in 3D gravity, where no such degeneracy is present. In contrast, in the 4D case the degeneracy forms an obstacle in relating the spacetime and phase space versions of the transformations we want to discuss, so we need to give a bit of intuition for the restriction we need to impose on the set of fields. Ultimately, it simply reflects the fact that the Hamiltonian evolution does not determine the connection field sufficiently, which must therefore be seen as having additional gauge freedom that is not apparent in the Lagrangian formulation. \\ 
    We will work with the following presentation of gravity as Einstein-Cartan-Holst theory\footnote{We will generically refer to these theories, regardless of their values for $\beta,\Lambda$, simply as \textit{tetrad gravity} throughout this work. This is meant to emphasize the \textit{methods} presented here over the specific \textit{theory} they are applied to, as well as evading issues of nomenclature.} with cosmological constant $\Lambda$:
    \begin{equation}\label{Action}
        S[\theta,\omega] = \int_M (\star+\beta)(\theta^2)\wedge F_\omega - \frac{2\Lambda}{4!} \theta^2\wedge\star\theta^2
    \end{equation}
    in which $\theta$ represents the tetrad, $\omega$ the Lorentzian (Lie algebra $\gfrak$) spin connection, $\beta = \frac{1}{\gamma}$ the Immirzi parameter\cite{Holst:1995pc} and the $\star$ indicates the internal Hodge dual.
    For simplicity, we will also use a shorthand for Lie algebra map appearing in the action,
    \begin{equation}
        (\theta^2)_{(\beta)} := (\star+\beta)(\theta^2),\quad \theta^2 := \theta\wedge\theta.
    \end{equation}
    We will use the covariant phase space approach\cite{harlowCovariantPhaseSpace2020,margalef-bentabolGeometricFormulationCovariant2021,Chandrasekaran:2021vyu,gomesUnifiedGeometricFramework2019,freidelExtendedCornerSymmetry2021,Fiorucci:2021pha,cattaneoReducedPhaseSpace2019}, although a canonical analysis along the same lines is equally possible. We will study the phase space of the theory on some fixed Cauchy slice $\Sigma$, and the off-shell (pre)phase space will be
    \begin{equation}\label{OffshelPhasespace}
        \Tilde{\ccal}_\Sigma = \Omega^1_{nd}(\Sigma,V)\times \acal(\Sigma,\gfrak)
    \end{equation}
    the product of nondegenerate pullbacks of tetrads (equivalently, maps into the fake tangent bundle $V$ whose image at all points of $\Sigma$ spans a 3D subspace of $V$) with the space of $\gfrak$-connections on the slice. 
    Out of this phase space we will select then the physical (pre)phase space of solutions of the Einstein-Cartan equations
    \begin{equation}
        \mathcal{E}=\delta\theta_I \wedge (G_\omega - \Lambda \star\theta^3)^I -d_\omega \theta^2_{(\beta)}\wedge\delta\omega \qquad
        G_\omega^I:= (F_\omega)^{IJ}_{(\beta)}\wedge\theta_J
    \end{equation}
    where we introduce the Einstein tensor $G_\omega$.\\
    In order to turn the set \ref{OffshelPhasespace} into a phase space, we equip it with the presymplectic potential coming from the boundary variation of the action \ref{Action}:
    \begin{equation}
        \delta S = \int_M \mathcal{E} + d \theta
    \end{equation}
    In general, one can split the variation of the action into a local bulk term $\mathcal{E}$ (the equations of motion), and a total derivative/total divergence $d\theta$ which gives rise to the symplectic potential of the theory.
    This is, in its integrated form
    \begin{equation}
        \Theta_\Sigma = \int_\Sigma (\theta^2)_\beta\wedge\delta\omega = -\int_\Sigma \theta_I \wedge \delta\omega_{(\beta)}^{IJ}\wedge\theta_J.
    \end{equation}
    The presymplectic form is produced by taking another variational derivative $\delta$, leading to the degenerate expression
    \begin{equation}
        \Tilde{\Omega}_\Sigma = \int_\Sigma \delta(\theta^2)_\beta\wedge\delta\omega = \int_\Sigma -\delta\theta_I \wedge \delta\omega^{IJ}_{(\beta)}\wedge\theta_J
    \end{equation}
    which has nontrivial kernel given by the vector fields\cite{cattaneoReducedPhaseSpace2019}
    \begin{equation}\label{DeltaBar}
        X_{\Bar{\Delta}} = \Bar{\Delta}\frac{\delta}{\delta\omega} \qquad \Bar{\Delta}_{(\beta)}^{IJ}\wedge\theta_J = 0
    \end{equation}
    which we will refer to as kernel vector fields or more simply as $\Bar{\Delta}$-vector fields.
    So, in fact, one needs to \say{fix a gauge} even on this kinematical level\footnote{Usually, gauge fixing is only necessary on-shell of the constraints in order to have a nondegenerate Poisson structure. Here, however, we already need to do this off-shell.}. The optimal way to do this has been found to be the so-called structural constraint on $\omega$\cite{cattaneoReducedPhaseSpace2019,canepaBoundaryStructureGeneral2021}. To state it, we first complete the (3-dimensional) image of the pullback of the tetrad to $\Sigma$, $\Tilde{\theta}$, to a 4-dimensional frame for the fake tangent bundle $V$. We do so with a fixed kinematical section $\nu,\delta\nu=0$ (chosen normalised to 1 in the internal space metric) which can be chosen functionally independent of the fields as long as $\Tilde{\theta}$ induces a nondegenerate metric on $\Sigma$.\footnote{For notational simplicity, for the majority of the paper we will work on the phase space level and drop the tilde as a restriction to the slide $\Sigma$ is implicit.}Then, the structural constraint can be stated as that $\nu\lor \Tilde{d}_\omega\Tilde{\theta}$ satisfies
    \begin{equation}
        \nu\lor \Tilde{d}_\omega\Tilde{\theta} = \tau\wedge\Tilde{\theta}
    \end{equation}
    for some $\tau \in \Omega^1(\Sigma,V)$.\footnote{$A\lor B$ denotes the internal wedge product of in $\Lambda^\bullet V$. } The left hand side has $3\cdot 6$ components acting as constraints, of which $3\cdot 4$ are alleviated by allowing for $\tau$ to be arbitrary. Therefore, overall $6$ constraints are imposed, fixing the kernel gauge precisely. \\
    This yields a unique seperation of any connection into a \say{reduced connection} and $\Bar{\Delta}$-part
    \begin{equation}
        \omega = \hat{\omega} + \Bar{\Delta}.
    \end{equation}
    Given a point $(\omega,\theta)$, one can then flow along the $\Bar{\Delta}$ orbits to a unique $(\hat{\omega},\theta)$, where the specific reduced connection $\hat{\omega}$ depends of course on the starting point.\\
    With respect to the coordinate split $(\theta,\omega)=(\theta,\hat{\omega},\Bar{\Delta})$, the gauge fixing is done simply by restricting to $\Bar{\Delta}=0$. Then, any observable on the phase space will be a functional of $\theta$ and $\hat{\omega}$ only, and extended by constancy to the $\Bar{\Delta}$-orbits. Naturally, the vector fields tangent to this gauge fixed phase space are then also of the form
    \begin{equation}
        X = X[\theta](\theta,\hat{\omega})\frac{\delta}{\delta\theta} + X[\hat{\omega}](\theta,\hat{\omega})\frac{\delta}{\delta\hat{\omega}},
    \end{equation}
    and the now nondegenerate symplectic form is
    \begin{equation}
        \Omega_\Sigma = -\int_\Sigma \delta\theta_I \wedge \delta\hat{\omega}^{IJ}_{(\beta)}\wedge\theta_J.
    \end{equation}
    The phase space itself is then the total space of an affine vector bundle
    \begin{equation}
        \ccal_\Sigma \rightarrow \Omega^1_{nd}(\Sigma,V)
    \end{equation}
    whose fibers are the spaces of reduced connections $\acal_{\text{red}}(\theta)$ satisfying the structural constraint.\\
    An obvious basepoint for the affine fibers is the torsion-free Levi-Civita connection $\gamma[\theta]$, so we can write all points in the fibre as
    \begin{equation}
        \hat{\omega} = \gamma[\theta] + \kappa,\; \kappa \in \Omega^1(\Sigma,\gfrak)\qquad \nu\lor q= \tau\wedge\Tilde{\theta}, q^I = \kappa^{IJ}{\wedge}\Tilde{\theta}_J
    \end{equation}
    We can interpret $\kappa$ as a contorsion 1-form which, by virtue of the structural constraint, is kinematically restricted to be of a certain form. This restriction, however, is only relevant off-shell or in the presence of matter.\\
    The general solution to the structural constraint can be given after decomposing $\kappa$ and $\tau$ with respect to the internal $\nu$:
    \begin{equation}
        \begin{gathered}
            \kappa = \kappa_\perp\lor\nu - \star(\kappa_\parallel\lor\nu) \qquad
            \kappa_\parallel^I = V^I_J \theta_\parallel^J  \\
            \tau^I = \tau^I_\parallel + \nu^I \tau_\perp \qquad
            \theta^I = \theta^I_\parallel + \nu^I \theta_\perp 
        \end{gathered}
    \end{equation}
    In this, the tangential $\kappa_\parallel^I,\tau_\parallel^I$ and normal $\kappa_\perp^I,\tau_\perp^I$ components to the spatial slice (seeing $\nu$ as the time direction) are subject to relations under the structural constraint. By expressing $\kappa_\parallel$ via some tensor $V^I_J$ in the tetrad basis, we can solve it for $\tau$
    \begin{equation}
        \begin{gathered}
            \tau_\perp = \epsilon_{ABCD}\theta_\parallel^A V^{BC}\nu^D \qquad \tau_\parallel = -\kappa_\perp
        \end{gathered}
    \end{equation}
    under the condition that
    \begin{equation}
        V_{(IJ)} = 0.
    \end{equation}
    In this, $\kappa$ is constrained to only have degrees of freedom encoded in some antisymmetric tensor $V_{IJ}$ orthogonal to $\nu$ (which makes it effectively a spatial 3x3 matrix with 3 degrees of freedom). With this restriction, one can then solve for $\tau$ in the form presented, meaning the structural constraint is fulfilled.\\
    This selects contorsions of the form
    \begin{equation}
        \kappa = \kappa_\perp\lor \nu + Q_n\lor\theta_\parallel \qquad Q_n^I \in \Omega^0 (\Sigma,V), Q_n\cdot\nu = 0
    \end{equation}
    We can get an even better idea of the form of these contorsions when using not a kinematical normal, but the \textit{adapted normal} $u_\Sigma$ to the slice, which is field-dependent:
    \begin{equation}\label{AdaptedNormal}
        U_I = V_I^\mu \mathbf{n}_\mu \qquad u_\Sigma=\frac{U}{||U||}
    \end{equation}
    It is defined only from the normal 1-form $\mathbf{n}$ to the slice $\Sigma$ and the tetrad (and its inverse $V$). In this decomposition, the solutions are all of the form
    \begin{equation}
        \kappa = i_{\hat{n}} (Q\lor\theta) \qquad Q \in \Omega^1(M,V) 
    \end{equation}
    where $\hat{n}$ is the vector field associated to $\mathbf{n}$ via the metric. Matching to the above, we have $\tilde{Q}=-\kappa_\perp,Q_n=Q_n$. Ostensibly, this means that the solutions break spacetime covariance, but only on the slice $\Sigma$, which does so by itself.
    In terms of the torsion $T^I = d_\omega\theta^I$, we can understand the restriction as the \textit{parallel part of the torsion, when expressed in the tetrad basis,}
    \begin{equation}
        T^I_\parallel \stackrel{\Sigma}{=}: A^I_J \wedge \theta^J
    \end{equation}
    \textit{being diagonal:}
    \begin{equation}
        \begin{gathered}
            A^{I}_J = R \delta^I_J \qquad R\in \Omega^1(\Sigma)
        \end{gathered}
    \end{equation}
    This is perhaps the simplest characterisation of the phase space of tetrad variable GR. We stress that this discussion shows an in-principle mismatch between the spacetime and phase space configuration variables of the theory: \textit{Only equivalence classes of connections constitute physical data even off-shell}\footnote{This is directly related to the notion of primary constraints in canonical analysis, which, in the covariant phase space formalism, are encoded in \textit{off-shell identities and gauge invariances on the configuration space}.}. On-shell in vacuum, the solution sets are the same, but as we are concerned with off-shell symmetries of the system, this distinction is crucial. In fact, we will see that the Kalb-Ramond translations we are seeking are sensitive to this mismatch.
    \\
    It is extremely useful to pick an arbitrary reference connection $\hat{\omega}_0$ and introduce the ADM momentum \say{aspect}\cite{freidelEdgeModesGravity2020a} 2-form as
    \begin{equation}
        p^I := -(\hat{\omega}-\hat{\omega}_0)^{IJ}_{(\beta)}\wedge\theta_J
    \end{equation}
    which, due to the structural constraint gauge fixing, is 1-to-1 with the reduced connections $\hat{\omega}$ for given $\omega_0$ as a basepoint.\\
    With this, we can rewrite the symplectic form as simply
    \begin{equation}
        \Omega_\Sigma = \int_\Sigma \delta\theta_I\wedge\delta p^I
    \end{equation}
    So upon a choice of (arbitrary!) basepoint section in the affine bundle $\ccal_\Sigma\rightarrow \Omega^1_{nd}$, we can realise the phase space as the cotangent bundle\cite{cattaneoReducedPhaseSpace2019,canepaBoundaryStructureGeneral2021}
    \begin{equation}
        \ccal_\Sigma \cong T^\ast \Omega^1_{nd}
    \end{equation}
    and the basepoint changes act on this cotangent bundle. Notice, however, that as $\omega_0$ is not a phase space coordinate, it does not transform under Lorentz transformations. In turn, $p$ does then not transform as a Lorentz vector. We will revisit this point in section \ref{DynRefFr}, where we offer an improved formulation of this variable.

\section{Phase space translation symmetries}
    We can now present the symmetry as it acts on the canonical phase space:
    With symbols to be introduced below, the vector field $Y_\phi$ is parameterised by a 0-form internal vector $\phi^I$ and acts as
    \begin{equation}
        Y_\phi[\theta] = d_{\hat{\omega}}\phi+ \mathbb{T}_\phi \qquad
        Y_\phi[\hat{\omega}] = \mathbb{F}_\phi - \Lambda \mathbb{L}_\phi
    \end{equation}
    which contains implicit expressions $\mathbb{T}_\phi$, $\mathbb{F}_\phi$ and $\mathbb{L}_\phi$ that we purposefully wish to distinguish from other quantities and which stem from the implicit definitions
    \begin{equation}
        \begin{gathered}
            \phi\lor d_\omega\theta = \mathbb{T}_\phi \wedge\theta\\
            \mathbb{F}_\phi\wedge\theta =  \phi\lor F_\omega,\\
             (\mathbb{L}_\phi)_{(\beta)} = \star(\phi\lor\theta).
        \end{gathered}
    \end{equation}
    Ki can be easily solved for as $\mathbb{L}_\phi = \frac{1+\beta\star}{1+\beta^2}(\phi\lor\theta)$, but we keep it implicit for the same reason as $\mathbb{F}_\phi$ - it rarely appears on its own, rather in the form found in the implicit definitions. Similarly, given the contorsion $\kappa$ from before, we have
    \begin{equation}
        \mathbb{T}_\phi^I = -\frac{1}{3}\kappa^{IJ}\phi_J
    \end{equation}
    whereas the expression for $\mathbb{F}_\phi$ involves the inverse triad in principle.\\
    In most calculations, one does not need the explicit forms of these objects. Instead, we can express their properties through the expressions
    \begin{equation}
        Y_\phi[\hat{\omega}]_{(\beta)}\wedge\theta= -( (F_\omega)_{(\beta)} - \Lambda \star\theta^2)\cdot \phi,
    \end{equation}
    and
    \begin{equation}
        Y_\phi[\theta\wedge\theta] = d_\omega(\phi\lor\theta),
    \end{equation}
    which are in fact sufficient to fully define $Y_\phi$.\\
    We first present a simple derivation of this transformation from a canonical Ansatz charge. After this, we discuss the relation to symmetries of the Lagrangian and highlight that there is a discrepancy between the above phase space transformation and the appropriate analogous symmetry of the Lagrangian. This explains the need to consider the phase space and properly account for the gauge fixing of the $\Bar{\Delta}$ vector fields.\\
    \subsection{Canonical derivation}
    By starting from the Einstein constraint as a function on phase space for a \textit{closed} slice $\Sigma$, we will be able to find relevant symplectic vector fields on the phase space.\\
    We have to first note that $G_\omega - \Lambda\star\theta^3 = 0$, unlike the Gauss constraint, is \textit{not} constant across the kernel foliation (not left constant by the $\Bar{\Delta}$-vector fields). Therefore, the functions
    \begin{equation}
        G_\omega \neq G_{\hat{\omega}}
    \end{equation}
    (where the latter is constant across orbits by definition) are not the same on the spacetime configuration space $\ccal_M$ or the prephase space $\Tilde{\ccal}_\Sigma$. This is not an accident: Together with the Gauss constriant, $G_\omega -\Lambda\star\theta^3= 0$ selects a \textit{unique representative} of each orbit in the kernel foliation. Therefore, on the on-shell phase space itself, there is no redundancy in the symplectic form. In a way, the Einstein constraint comes \say{equipped} with a gauge fixing for the $\Bar{\Delta}$-transformations. However, when writing the offshell phase space using the gauge fixing of the structural constraint, the restriction to the onshell phase space is correctly performed by setting $G_{\hat{\omega}} = \Lambda\star\theta^3$.\cite{cattaneoReducedPhaseSpace2019,canepaBoundaryStructureGeneral2021}\\
    We can take the variation of the Einstein tensor directly, keeping in mind that the variations are constrained to preserve the structural constraint.
    \begin{equation}
        \phi_I\delta G_{\hat{\omega}}^I = (\phi\lor\theta)_{(\beta)}\wedge d_{\hat{\omega}}\delta\hat{\omega} -\delta\theta_I (F_{\hat{\omega}})_{(\beta)}^{IJ}\phi_J
    \end{equation}
    For the cosmological constant contribution, we also have
    \begin{equation}
        \phi_I \delta(\star\theta^3)^I = \frac{1}{3}\delta\theta_I (\star\theta^2)^{IJ}\phi_J 
    \end{equation}
    We can partially integrate this and for now neglect the boundary term to study which vector field will actually be Hamiltonian coming from the constraint.
    We therefore suppose $\delta\phi =0$ and
    \begin{equation}
        \begin{aligned}
            -I_Y\Omega &= \int_\Sigma\delta(-\phi\cdot (G_{\hat{\omega}}-\Lambda \star\theta^3))\\
            &=-\int (d_{\hat{\omega}}\phi\wedge\theta+ \phi\lor d_{\hat{\omega}}\theta)\wedge\delta\hat{\omega}_{(\beta)}\\
            &-\delta\theta_I (F_{\hat{\omega}})_{(\beta)}^{IJ}\phi_J
            -\delta\theta_I \Lambda\star(\phi\lor\theta)^{IJ}\theta_J \\
            &=\int (d_{\hat{\omega}}\phi+ \mathbb{T}_\phi)_I\wedge\delta\hat{\omega}_{(\beta)}^{IJ}\wedge\theta_J
            +\delta\theta_I ((F_{\hat{\omega}})_{(\beta)}^{IJ}\phi_J 
            +
            \Lambda\star(\phi\lor\theta)^{IJ}\theta_J 
            )
        \end{aligned}
    \end{equation}
    Using equation \ref{Contraction} from appendix \ref{SympVFDetail}, one can see that it generates the Hamiltonian vector field
    \begin{equation}
        Y_\phi[\theta] = d_{\hat{\omega}}\phi+ \mathbb{T}_\phi \qquad
        Y_\phi[\hat{\omega}] = \mathbb{F}_\phi - \Lambda \mathbb{L}_\phi.
    \end{equation}
    So what is correctly shifted in 4-dimensional gravity by a derivative is not the tetrad, but the gravitational flux $\theta^2$, showing that this symmetry is a remnant from BF theory\footnote{In 4D BF theory, the 2-form Kalb-Ramond field $B$ is shifted as $B\mapsto B + d_\omega\mu$ with a Lie algebra-valued 1-form $\mu$.} after imposition of the simplicity constraints to arrive at gravity.
    Using the adapted normal (\ref{AdaptedNormal}) for decompositions (see Appendix \ref{App:Ashtekar} for details on convention), we can express the action of this vector field on the triad and the normal using the inverse triad $\hat{v}$ and the extrinsic curvature $K = d_\omega u= K_\alpha dx^\alpha$:
    \begin{equation}
        \begin{gathered}
            Y_\phi[u]_I = - \hat{v}^\alpha_I (\partial_\alpha \phi_\perp - \frac{2}{3}K_\alpha^I(\phi_\parallel)_I - \frac{1}{3}(K_{\gamma[\theta]})^I_\alpha (\phi_\parallel)_I)\\
            \widetilde{Y_\phi[\Tilde{e}]}_\parallel = d_\Gamma\phi_\parallel + \frac{2}{3}\phi_\perp K - \frac{1}{3} [\Gamma_\parallel - \gamma[\theta]_\parallel]\times\phi_\parallel + \frac{1}{3}\phi_\perp K_{\gamma[\theta]}
        \end{gathered}
    \end{equation}
    In particular, we can see the presence of a $d_\Gamma\phi_\parallel$ term for the triad, which is the usual Kalb-Ramond-type translation piece, as also used for the translations on the phase space of Loop Quantum Gravity\cite{freidelKinematicalGravitationalCharge2020}. However less straightforwardly, the timelike translations instead mix the triad and the extrinsic curvature. Particularly, on-shell of the Gauss constraint,
    \begin{equation}
        \begin{gathered}
            Y_\phi[u]_I = - \hat{v}^\alpha_I (\partial_\alpha \phi_\perp  - (K_{\gamma[\theta]})^I_\alpha (\phi_\parallel)_I)\\
            \widetilde{Y_\phi[\Tilde{e}]}_\parallel = d_{\gamma[\theta]_\parallel}\phi_\parallel +  \phi_\perp K_{\gamma[\theta]}
        \end{gathered}
    \end{equation}
    we only have the Levi-Civita connection and the extrinsic curvature is that of the Cauchy surface in the dynamical metric. In particular, for moment-of-time symmetric slices $K_{\gamma[\theta]}=0 $, the time translations are trivial on the triad - as expected, as this also implies vanishing of the ADM momentum. However still the timelike normal $u$ is affected nontrivially.\\
    This way of constructing a symmetry has an analogy in other gauge theories of differential forms, where the equations of motion can be pulled back to spatial slices to give constraints. Characteristically, the canonical generators created from these constraints take the form
    \begin{equation}
        P_\phi = \int_\Sigma C_\phi + \oint_{\partial\Sigma}q_\phi
    \end{equation}
    where the bulk constraint vanishes on-shell $C_\phi\approx 0$, so that at most a corner charge $q_\phi$ remains non-zero. The bulk piece then characterises the gauge structure of the theory, whereas the corner charge gives an additional set of true symmetries labeling the theory when spatial boundaries are present.\\
    \subsection{Contrast to symmetries of the Lagrangian}
    The presence of such constraints comes, by Noether's second theorem in a 1-to-1 way, with gauge symmetries of the theory. However, we need to be careful in invoking said theorem as we have not made use of Lagrangians so far. 
    In fact, the vector field we presented is \textit{not} the phase space pushforward of a transformation acting on the spacetime fields $\theta,\omega$ appearing in the Lagrangian. 
    There is still an analogue of it, which has been previously studied by  Montesinos et. al\cite{montesinosReformulationSymmetriesFirstorder2017}, where by use of Noether identities, one obtains a symmetry of the Lagrangian that looks very similar. For this, we simply take derivatives of the Einstein tensor (this time in spacetime),
    \begin{equation}
        d_\omega G_\omega^I = (F_\omega)_{(\beta)}^{IJ}\wedge d_\omega\theta_J
    \end{equation}
    and weigh it by an arbitrary internal 4-vector $\phi^I$ to arrive at the Noether identity
    \begin{equation}\label{NoetherId}
        d(\phi_I G_\omega^I) = d_\omega\phi_I \wedge G_\omega^I +  (\phi\lor d_\omega\theta)\wedge (F_\omega)_{(\beta)}.
    \end{equation}
    For the cosmological constant part, we will also need
    \begin{equation}
        d(\phi_I \star(\theta^3)^I) = d_\omega\phi_I\wedge \star(\theta^3)^I + (\phi\lor d_\omega\theta)\wedge\star\theta^2.
    \end{equation}
    This can be used to find vector fields which satisfy the characteristic relation of local symmetries,
    \begin{equation}
        I_X\mathcal{E} = d C.
    \end{equation}
    Here, $C$ is a codimension 1 form that vanishes on-shell of the equations of motion, and is referred to as the \textit{constraint form}.
    For this, we need to rewrite the torsion term. There are essentially two options for this, which we can combine in a general way. 
    First, note that the general contraction is
    \begin{equation}
        I_X \mathcal{E} = X_\phi[\theta]_I\wedge (G_\omega-\Lambda\star\theta
        ^3)^I- d_\omega \theta^2_{(\beta)}\wedge X_\phi[\omega]
    \end{equation}
    The first of the two ways is to rewrite
    \begin{equation}
        \phi\lor d_\omega\theta = \mathbb{T}_\phi\wedge\theta.
    \end{equation}
    The second option is to use $\mathbb{F}_\phi$ such that 
    \begin{equation}
        (\mathbb{F}_\phi)^{IJ}_{(\beta)}\wedge\theta_J = - (F_\omega)^{IJ}_{(\beta)}\phi_J,
    \end{equation}
    which reproduces the torsion term in the contraction. Also useful is $\mathbb{L}_\phi$ for the cosmological constant terms.
    We can then see by mixing and matching these ingredients that the family of vector fields for $s\in \rbb$
    \begin{equation}\label{STVectorFields}
        X_{\phi,s}[\theta] = d_\omega\phi + s \mathbb{T}_\phi \qquad 
        X_{\phi,s}[\omega] = (1-s)\mathbb{F}_\phi + \Lambda (1-\frac{s}{3}) \mathbb{L}_\phi
    \end{equation}
    are all local symmetries:
    \begin{equation}
        I_X\mathcal{E} = d\left(\phi_I(G_\omega - \Lambda\star\theta^3)^I\right)
    \end{equation}
    \begin{equation}
        X_{\phi,s}[L] = d\left( 
\phi_I (  G_\omega -\Lambda \star\theta^3)^I + \theta^2_{(\beta)}\wedge X_\phi[\omega] \right) \qquad \forall s,\phi
    \end{equation}
    These are associated to the Noether currents
    \begin{equation}
        j^\phi_s = \phi_I (G_\omega - \Lambda\star\theta^3)^I
    \end{equation}
    which all vanish on-shell.
    However, we can already see that the pushforward of this vector field to the phase space $\ccal_\Sigma$ will have issues: By analysing the symplectic form in detail, as we do in Appendix \ref{SympVFDetail}, we can find criteria for a vector field to be symplectic. 
    Then the vector fields here will not be symplectic in general. \\
    This means, in particular, that the symmetry of the Lagrangian can not be canonically represented. In a potential quantisation, this is unfortunate on a technical level and it is unclear how to realise such the action of a symmetry on the system through (unitary) operators.
    That the two transformations do not agree is perhaps more than an unlucky coincidence: It is known from careful BV-BFV studies\cite{canepaBoundaryStructureGeneral2021,canepaGeneralRelativityAKSZ2021a} that the spacetime formulation (BV) of tetrad gravity and the phase space (BFV) formulation, while seperately equivalent to the Einstein-Hilbert formulations, are \textit{not} equivalent in a more strict sense (BV-BFV).  What this means is the following: The Hamiltonian evolution structure of ECH and its Lagrangian equations of motion are \say{incompatible} in the sense that there is a discrepancy in configuration spaces:
    \begin{equation}
        \begin{gathered}
            \ccal_M^{Lagr}= \Omega^1_{nd}(M,V)\times \acal(M,\gfrak)\\
            \ccal_M^{Ham}= \Omega^1_{nd}(M,V)\times \acal_{red}(M,\gfrak)
        \end{gathered}
    \end{equation}
    The \say{reduced} connections $\acal_{red}$ are simply the ones satisfying the structural constraint with respect to a given local foliation of the spacetime $M$ into spatial slices. So, while the Lagrangian equations of motion are defined for all possible spin connections, the Hamiltonian evolution is only sensible for their equivalence class under the $\Bar{\Delta}$-transformations. 
    BV-BFV compatibility states now, in essence, that taking a covariant spacetime dynamics, restricting it to an initial slice $\Sigma$, and then evolving canonically on a cylinder $\Sigma\times \rbb$ returns one to the same spacetime dynamics. Clearly, this is not the case in ECH as presented here.\\
    This can in principle be remedied by restricting the spacetime configurations $\omega$ to $\hat{\omega}$, but this cannot be done in a covariant way. Overall, we can see that it is no surprise that the spacetime covariant symmetries $X$ do not descend necessarily to the phase space. Therefore, as long as we want to perform phase space studies, we will want to stick to the vector fields $Y$, as these are the ones canonically represented. \\
    We can actually make this point clear by studying the time evolution on the phase space: by splitting the equations of motion
    \begin{equation}
        G_\omega = 0 = d_\omega\theta\wedge\theta
    \end{equation}
    into their horizontal (constraint) and vertical (evolution) parts, we can find the evolution equations of the canonical variables of the slice. We find, in particular, that they entail
    \begin{equation}
        \begin{gathered}
            \mathcal{L}_{\hat{n}}\Tilde{\theta} \approx \Tilde{d}_{\omega}\theta_n + \tau_{\theta_n} - \omega_n \cdot \Tilde{\theta}\\
        \mathcal{L}_{\hat{n}}\Tilde{\omega} \approx 
        \Tilde{d}_{\omega}\omega_n + Y_{\theta_n}[\Tilde{\omega}] + \Bar{\Delta}
        \end{gathered}
    \end{equation}
    where $\omega_n=i_{\hat{n}}\omega,\theta_n=i_{\hat{n}}\theta $ and the $\Bar{\Delta}$ piece is \textit{arbitrary} and precisely of the type that is removed by the kernel quotient $\tilde{\omega}\mapsto [\tilde{\omega}]\ni \hat{\omega} $ - its presence states that the time evolution of $\Tilde{\omega}$ under this equation is gauge invariant under the kernel foliation, and so, for the reduced connection, well defined. In principle, then, one must fix this gauge freedom in order to have well-posed time evolution. The $Y$-piece is precisely the vector field from the previous section, so in practice we have that
    \begin{equation}
        \mathcal{L}_{\hat{n}} \approx C_{\omega_n} + Y_{\theta_n}
    \end{equation}
    or in other words, time evolution in tetrad gravity is pure gauge and a combination of Lorentz transformations and the new, nonlinear Kalb-Ramond transformations.\\
    If, instead, we decompose the action of general diffeomorphisms on the spacetime fields in a similar way, by appealing to the split of diffeomorphisms into Lorentz transformations and covariant diffeomorphisms
    \begin{equation}
        \begin{gathered}
            \mathcal{L}_\xi \theta = d_\omega i_\xi\theta - i_\xi\omega\cdot \theta + i_\xi d_\omega\theta\\
            \mathcal{L}_\xi \omega = i_\xi F_\omega + d_\omega(i_\xi \omega),
        \end{gathered}
    \end{equation}
    then regular diffeos can then be rewritten as \textit{yet different} field-dependent translations
    \begin{equation}
        Z_\phi[\theta]^I = d_\omega\phi^I + \phi^J i_{\hat{V}_J}d_\omega\theta^I \qquad Z_\phi[\omega]^{IJ} = \phi^K i_{\hat{V}_K}F_\omega^{IJ}
    \end{equation}
    together with Lorentz transformations $C_\alpha$
    \begin{equation}
        \mathcal{L}_\xi = C_{i_\xi \omega} + Z_{i_\xi \theta}.
    \end{equation}
    While these other translations $Z_\phi$ might superficially look similar to the $Y_\phi$ ones, the latter are different due to the contraction pattern of $\phi$. They are also not symmetries of the Lagrangian like $X_{\phi,s}$.\\
    We can therefore see that unlike in the case of 3D gravity, there are a multitude of translations depending on the motivation. The presence of canonical generators and compatibility with time evolution singles out the transformation $Y_\phi$ as the optimal choice for generators, leaving a schism between the Lagrangian and Hamiltonian formulation. Nevertheless, these transformations are genuine gauge symmetries of the Hamiltonian theory and can be used to rewrite diffeomorphisms as effective, field-dependent transformations like in 3D gravity.\\

\subsection{Bulk Poisson brackets}
    We now calculate the commutation relations of these new, nonlinear Kalb-Ramond translations and find (as expected) that they close with structure functions.  
    On the phase space of a closed slice $\Sigma$, the vector field $Y_\phi$ is generated by the Einstein constraint
    \begin{equation}
        P_\phi = -\int_\Sigma \phi_I (G_\omega-\Lambda \star\theta^3)^I.
    \end{equation}
    In this, and from now on, we mostly drop the hat $\hat{\omega}$ from the reduced connections and work entirely on the slice, so that we can also suppress pullbacks.\\
    We can directly calculate its Poisson brackets with itself via the action of the vector field on the charge: 
    \begin{align}
        Y_\phi[P_{\Tilde{\phi}}] &=
        -\int_\Sigma (\Tilde{\phi}\lor\theta)_{(\beta)} d_\omega(Y_\phi[\omega]) + \Tilde{\phi}_I Y_\phi[\theta]_J ((F_\omega)_{(\beta)} - \Lambda
       (\star\theta^2))^{IJ}\\
       &=
       -\int_\Sigma  \Tilde{\phi}_I Y_\phi[\theta]_J ((F_\omega)_{(\beta)} - \Lambda
       (\star\theta^2))^{IJ}
       + d_\omega(\Tilde{\phi}\lor\theta)_{(\beta)} (Y_\phi[\omega]) \\
       &=
       -\int_\Sigma  \Tilde{\phi}_I Y_\phi[\theta]_J ((F_\omega)_{(\beta)} - \Lambda
       (\star\theta^2))^{IJ}
       +(Y_{\Tilde{\phi}}[\theta]\wedge\theta) \cdot(Y_\phi[\omega])_{(\beta)} \\
       &=
       -\int_\Sigma  \Tilde{\phi}_I Y_\phi[\theta]_J ((F_\omega)_{(\beta)} - \Lambda
       (\star\theta^2))^{IJ}\\
       &\;-\int_\Sigma Y_{\Tilde{\phi}}[\theta]_I ((F_\omega)_{(\beta)}^{IJ}\phi_J + \Lambda \star(\phi\lor\theta)^{IJ}\theta_J)\\
       &= 
       -\int_\Sigma  \Tilde{\phi}_I Y_\phi[\theta]_J (F_\omega)_{(\beta)}^{IJ}
       +Y_{\Tilde{\phi}}[\theta]_I (F_\omega)_{(\beta)}^{IJ}\phi_J \\
       &\;+\int_\Sigma \Lambda Y_{\Tilde{\phi}}[\theta]_I \star(\phi\lor\theta)^{IJ}\theta_J
       + \Tilde{\phi}_I Y_\phi[\theta]_J\Lambda
       (\star\theta^2)^{IJ}
       \\
       &=
       -\int_\Sigma  (\Tilde{\phi}\lor Y_\phi[\theta]+Y_{\Tilde{\phi}}[\theta]\lor\phi) \cdot(F_\omega)_{(\beta)} \\
       &\;+\int_\Sigma \Lambda (\Tilde{\phi}\lor Y_\phi[\theta]+Y_{\Tilde{\phi}}[\theta]\lor\phi) \cdot
       (\star\theta^2)\\
       &= \{P_\phi, P_{\Tilde{\phi}}\}=P_{\Phi(\phi,\Tilde{\phi})}.
    \end{align}
    In this, we integrated by parts in the first line, used the identities of $\mathbb{T}_\phi$ in the second, and the ones of $\mathbb{F}_\phi$, $\mathbb{L}_\phi$ in the third. The charges therefore close with structure functions $\Phi$ which are implicitly defined via
     \begin{equation}
        \Phi(\phi,\Tilde{\phi})\lor \theta = \Tilde{\phi}\lor Y_\phi[\theta]+Y_{\Tilde{\phi}}[\theta]\lor\phi.
    \end{equation}
    This is clearly \textit{not} a simple translation algebra like in 3D gravity, but this is of no surprise as 4D gravity is much richer.
    On-shell in vacuo, the above relation simplifies to \begin{equation}\label{OnShellStructureFunctions}
        \Phi(\phi,\Tilde{\phi})\lor \theta =d_{\omega}(\Tilde{\phi}\lor \phi)= d_{\gamma[\theta]}(\Tilde{\phi}\lor \phi)
    \end{equation}
    involving only the Levi-Civita covariant derivative. This allows us to see simplification of the Poisson relations when torsion is absent. By using the expression of $\mathbb{T}_\phi=-\frac{1}{3}\kappa\cdot \phi$ through the contorsion $\kappa$, we can rewrite
    \begin{equation}
        \Tilde{\phi}\lor Y_\phi[\theta]+Y_{\Tilde{\phi}}[\theta]\lor\phi = d_\omega(\Tilde{\phi}\lor\phi) + [\Tilde{\phi}\lor\phi,\frac{1}{3}\kappa]
    \end{equation}
    which gives us the more explicit expression of the Poisson bracket as
    \begin{equation}
        \begin{aligned}
            \{P_\phi, P_{\Tilde{\phi}}\} &= 
            \int_\Sigma \Lambda \star\theta^2\wedge \left(d_\omega(\Tilde{\phi}\lor\phi) + [\Tilde{\phi}\lor\phi,\frac{1}{3}\kappa]\right)\\
            -&\int_\Sigma (F_\omega)_{(\beta)}\wedge \left(d_\omega(\Tilde{\phi}\lor\phi) + [\Tilde{\phi}\lor\phi,\frac{1}{3}\kappa]\right)
        \end{aligned}
    \end{equation}
    So technically, if we defined the functions
    \begin{equation}
        R_\alpha := -\int_\Sigma ((F_\omega)_{(\beta)}-\Lambda\star\theta^2)\wedge \left(d_\omega\alpha + [\alpha,\frac{1}{3}\kappa]\right)
    \end{equation}
    we would have $\{P_\phi, P_{\Tilde{\phi}}\} = R_{\Tilde{\phi}\lor\phi}$. However, when $\kappa=0$, as happens in vacuum on-shell, this simplifies to familiar expressions due to the Bianchi identity $d_\omega F_\omega=0$:
    \begin{equation}
        R_\alpha \stackrel{\kappa=0}{=} \Lambda\int_\Sigma \theta^2_{(\beta)}\wedge d_\omega\Bar{\alpha} - \int_\Sigma d((F_\omega)_{(\beta)}\cdot\alpha) 
    \end{equation}
    where $\Bar{\alpha} := \frac{1+\beta\star}{1+\beta^2}\alpha$. This shows that for closed slices $\partial\Sigma=\emptyset$, the curvature term drops out and the only noncommutativity is due to the cosmological constant term:
    \begin{equation}
        \{P_\phi, P_{\Tilde{\phi}}\} \stackrel{\kappa=0}{=} \Lambda \, C_{\frac{1+\beta\star}{1+\beta^2}(\Tilde{\phi}\lor\phi)} = \Lambda \int_\Sigma \star\theta^2\wedge d_\omega(\tilde{\phi}\lor\phi)
    \end{equation}
    So we an see that when we are on the torsion-free subspace, two translations give rise to a Lorentz transformation. As a special case, the translations become commutative when the cosmological constant vanishes. We can therefore see the complications in the structure functions as \textit{not} due to the presence of curvature, but instead due to torsion.
    \\
    We also report that, by similar calculation, the Poisson brackets with the Lorentz charges
    \begin{equation}
        C_\alpha = \int_\Sigma \theta^2_{(\beta)}\wedge d_\omega\alpha
    \end{equation}
    is much simpler:
    \begin{equation}
        \{C_\alpha,P_\phi\} = P_{\alpha\cdot\phi}
    \end{equation}
    which is the action of Lorentz transformations on translations. Therefore, up to the above structure functions, the gauge algebra present in 4D tetrad gravity is very similar to a Poincare/deSitter one, mirroring the spacetime transformation result found by Montesinos et al.\cite{montesinosReformulationSymmetriesFirstorder2017}
    In fact, once we go onto the torsion-free subspace, we recover a Poincare or deSitter type algebra directly.

\subsection{Path systems and integrability on corners}
    Having motivated our choice for the covariant translations, we will proceed with a construction of the canonical charges. For this, we will work with the vector field $Y_\phi$ as it is the one with a candidate Hamiltonian generator which, as required of gauge generators, vanishes in the bulk on-shell.
    As stated before, $Y_\phi$ is the Hamiltonian vector field on $\ccal_\Sigma$ associated to the generator given by the Einstein constraint when the slice has no corner, and the parameter $\phi$ is field-independent:
    \begin{equation}
        \begin{gathered}
            I_{Y_\phi}\Omega_\Sigma + \delta P^{(0)}_\phi = 0\\
            P^{(0)}_\phi = -\int_\Sigma (\phi\lor\theta)\wedge ((F_\omega)_{(\beta)}-\Lambda \star\theta^2)
        \end{gathered}
    \end{equation}
    We are now interested in the case with corner, and particularly in which corner terms are added to the generator. We see that with corners, 
    \begin{equation}
         \delta P^{(0)}_{\phi} - P^{(0)}_{\delta\phi} + I_{Y_\phi}\Omega =
        - \oint_{\partial\Sigma} (\phi\lor\theta)_{(\beta)}{\wedge}  \delta\omega =   - \oint_{\partial\Sigma} \phi_I \alpha^I.
    \end{equation}
    \begin{equation}
        \alpha^I := -\delta\omega^{IJ}_{(\beta)}\wedge\theta_J
    \end{equation}
    This immediately seems to require an extension of the charge. Let us take inspiration from the fact that tetrad gravity can be written as a constrained BF theory\cite{celadaBFGravity2016a}, starting from 
    \begin{equation}
        L = B\wedge F_\omega - \frac{\Lambda}{2}B\wedge\Psi(B)
    \end{equation}
    where $\Psi$ is some constant Lie algebra homomorphism.
    The bulk charge resembles that of the 1-form Kalb-Ramond translations of BF theory, which has full form
    \begin{equation}\label{BFKRC}
        K_\mu = -\int_\Sigma \mu\wedge (F_\omega-\Lambda \Psi(B)) - \oint_{\partial\Sigma} \mu \wedge \omega
    \end{equation}
    where we would need to equate $\mu = (\phi\lor\theta)_{(\beta)} $.
    Then, it appears the charges we should try are given by
    \begin{equation}
        P_\phi = -\int_\Sigma (\phi\lor\theta)_{(\beta)}\wedge F_{\hat{\omega}} -\oint_{\partial\Sigma}(\phi\lor\theta)_{(\beta)}\wedge \hat{\omega}.
    \end{equation}
    This improves the situation mildly, but leaves us with another nonintegrable term:
    \begin{equation}
        \delta P_{\phi} - P_{\delta\phi} + I_{Y_\phi}\Omega =  \oint_{\partial\Sigma} \phi_I \omega^{IJ}_{(\beta)}\wedge\delta\theta_J =: \oint_{\partial\Sigma} \phi_I \gamma^I.
    \end{equation}
    So quite directly, we can give charges for Dirichlet boundary conditions on either $\hat{\omega}$ (no corner term) or $\theta$ (the corner term above) this way. However, this is not satisfactory in general as this charge only pertains to a subsector of gravity.\\
    In principle, $\alpha$ is not even closed, which is the origin of the integrability issues. Strict integrability therefore only holds when
    \begin{equation}\label{IntegrabilityCondition}
        \delta\alpha^I = \delta\omega^{IJ}_{(\beta)}\wedge\delta\theta_J \stackrel{\partial\Sigma}{=} 0.
    \end{equation}
    This imposes conditions on allowed (field-dependent) values of $\phi,\alpha$ in translations and Lorentz transformations. While the most general conditions seem out of reach, for $\hat{\omega}$-Dirichlet conditions they are
    \begin{equation}
            ((F_{\omega})_{(\beta)} -\Lambda\star\theta^2)\cdot\phi \stackrel{\partial\Sigma}{=} 0 
    \end{equation}
    whereas for $\theta$-Dirichlet they are
    \begin{equation}
        d_\omega(\phi\lor\theta)\stackrel{\partial\Sigma}{=}0 \quad \text{or on-shell:  } d_\omega\phi\stackrel{\partial\Sigma}{=}0.
    \end{equation}
    Clearly, these are quite restrictive unless specific boundary conditions are chosen, for example in the former situation that $\hat{\omega}$ is of constant curvature $\Lambda$, in which case all $\phi$ are admissible, but the corner charges vanish regardless. In the latter, however, restrictions of the values of $\phi$ seem unavoidable. \\ 
    In the $\hat{\omega}$-Dirichlet case, the Poisson commutation relations are almost unchanged from the previous section. The action of Lorentz transformations on translations is the same, and for translations themselves, bulk and corner terms cancel to give
    \begin{equation}
        \{P_\phi, P_{\Tilde{\phi}}\} \stackrel{\kappa=0}{=} -\Lambda \int_\Sigma d_\omega\star\theta^2\cdot(\tilde{\phi}\lor\phi) \stackrel{\kappa=0}{=} 0
    \end{equation}
    so that in fact, it also takes the form of a Lorentz bulk term.
    So in the end, when $\Lambda=0$, with these boundary conditions we again find commuting translations, and with $\Lambda\neq 0$ we find a resulting bulk Lorentz transformation, which is trivial on-shell. So effectively, for \textit{any} choice of $\Lambda$, if $\hat{\omega}$-Dirichlet boundary conditions hold, translations are commuting on-shell of the Gauss constraint.\\
    In the $\theta$-Dirichlet case we find instead, due to the new corner term,
    \begin{equation}
        \{P_\phi,P_{\Tilde{\phi}}\} \stackrel{\kappa=0}{=} -\Lambda \int_\Sigma d_\omega\star\theta^2\cdot(\tilde{\phi}\lor\phi) 
        - \oint_{\partial\Sigma}(\tilde{\phi}\lor\phi)_{(\beta)}\cdot F_\omega
    \end{equation}
    So, we find that the corner translations become noncommutative when the tetrad is fixed. However, we must keep in mind the constancy condition $d_\omega \phi = 0$ coming from the boundary condition. On-shell of the Gauss constraint, this corner term is effectively central.\\
    Interestingly, we also find 
    \begin{equation}
        \{C_\alpha,P_\phi\} = P_{\alpha\cdot\phi}-\oint_{\partial\Sigma} (\phi\lor\theta)_{(\beta)}\wedge d\alpha
    \end{equation}
    by using that $\theta$ must also be Lorentz-preserved in these boundary conditions. This is quite similar to the situation in 3D gravity, where
    \begin{equation}
        \{C_\alpha,P_\phi\} = P_{\alpha\cdot\phi}-\oint_{\partial\Sigma} \star\phi\cdot d\alpha
    \end{equation}
    also shows a similar central extension on corners. Unlike the 3D case, though, there now is also an extension for the translations.\\
     So in fact, for these boundary conditions on $\theta$, the true corner symmetry is 
    \begin{equation}
        \text{Stab}(\theta,SO(1,3)^{\partial\Sigma})\ltimes (H)^{\partial\Sigma}_{d_{\gamma[\theta]}\phi=0}
    \end{equation}
    where $H=\rbb^{1,3}$ with the Poisson bracket given above in terms of the curvature. This is as close to a Poincare algebras as we can get.\footnote{We note that this is the set of on-shell, boundary condition preserving gauge symmetries. These are to be contrasted with gauge transformations that have either vanishing generator (in which case they are a redundancy) and transformations not preserving the boundary condition (in which case they are not canonically represented on a given single phase space for a closed system).} This is to be contrasted with the $\hat{\omega}$-Dirichlet case, where the corner symmetry is, due to vanishing charges for the translations,
    \begin{equation}
        \text{Stab}(\hat{\omega},SO(1,3)^{\partial\Sigma}).
    \end{equation}
    It should be clear that both of these boundary conditions are quite restrictive and requiring strict integrability is limiting.\\
    Still, we can integrate the $-I_Y\Omega$ above along a phase space curve. If we choose a curve at constant $\theta$, e.g. a linear
    \begin{equation}
        \gamma(t) = (\omega_0 + t(\hat{\omega}-\omega_0), \theta)\qquad t\in[0,1]
    \end{equation}
    then the resulting integrals are
    \begin{equation}
        \int_\gamma \alpha^I = p^I :=  (\hat{\omega}-\omega_0)_{(\beta)}^{IJ}\wedge\theta_J 
    \end{equation}
    so just the ADM momentum aspect we defined earlier.
    The resulting function is then
    \begin{equation}
        \begin{aligned}
            -\int_\gamma I_{Y_\phi}\Omega_\Sigma &=
        \int_\Sigma (\phi\lor\theta)_{(\beta)}\wedge F_{\hat{\omega}} - (\phi\lor\theta)_{(\beta)}\wedge F_{\omega_0} - \oint_{\partial\Sigma} \phi_I p^I\\
        &= P_\phi(\hat{\omega},\theta)-P_\phi(\omega_0,\theta)
        \end{aligned}    
    \end{equation}
    where 
    \begin{equation}
        P_\phi(\hat{\omega},\theta) := 
        \int_\Sigma (\phi\lor\theta)_{(\beta)}\wedge F_{\hat{\omega}}
        - \oint_{\partial\Sigma} (\phi\lor\theta)_{(\beta)}\wedge \hat{\omega}
    \end{equation}
    This is the same function we originally proposed from the form of the BF charges. So, in a restricted sense, it is possible to find charges: One needs to define them along some path in phase space which does not change the value of $\theta$, and with respect to some reference connection. These are therefore more like \say{relative} translation charges. This is already common in gravity - the Iyer-Wald diffeomorphism charges in most of the literature are similarly nonintegrable and require an offset, at the very least for finiteness\cite{iyerPropertiesNoetherCharge1994a,Chandrasekaran:2021vyu,harlowCovariantPhaseSpace2020,Fiorucci:2021pha}. The obvious questions are what effects the choice of path system has in this, as well as which subset of phase space features integrability of the charges, so independence of the path.\\
    For this, let us start over with an a priori fixed path system $\{\gamma(\theta,\omega)\}$. Then, we can equally well \textit{define} $p$ as
    \begin{equation}
        p^I_\gamma (\theta,\hat{\omega}) := \int_{\gamma(\theta,\hat{\omega})} \alpha^I.
    \end{equation}
    For the linear path system above, we get back the already-used definition.
    We can then also define again the integrated charge with respect to this path system, given by 
    \begin{equation}
        P_{\phi;\gamma}(\theta,\hat{\omega}) := \int_{\gamma(\theta,\hat{\omega})} -I_{Y_\phi}\Omega_\Sigma.
    \end{equation}
    How should such a path system be chosen? The linear one is appealing for its simplicity. Yet still, we might want to be more thoughtful here. In particular, we should ask for constancy of the function along the flow lines of the vector field $Y_{\phi}$. This is necessary for it to be \say{conserved along its flow}, if it did have a flow. \\
    However, the calculations below show that quite luckily, the thus defined charges with respect to the linear path system are conserved along their flow in the sense that the Poisson bracket $\{P_\phi,P_\phi\}=0$ vanishes always. So, this choice of path system seems warranted. So overall, we could use $P_\phi$ with the implicit caveat that the charges should be used as relative ones.\\
    We can also take the viewpoint that the nonintegrability represents some flux of data leaving the system. In this sense, the problem becomes analogous to that of diffeomorphism charges: We need to find an appropriate split between integrable ($Q_\phi$) and nonintegrable ($\fcal_\phi$) corner pieces\cite{barnichSymmetriesAsymptoticallyFlat2010,freidelExtendedCornerSymmetry2021,Chandrasekaran:2021vyu,ciambelliIsolatedSurfacesSymmetries2021a,Fiorucci:2021pha,freidelWeylBMSGroup2021a}
    \begin{equation}
        -I_{Y_\phi}\Omega = \delta P^{(0)}_\phi + \delta Q_\phi - \fcal_\phi.
    \end{equation}
    With such a split made, one can then find an adapted charge bracket
    \begin{equation}
        \{Q_\phi,Q_{\Tilde{\phi}} \}_c
        := Y_\phi[Q_{\Tilde{\phi}}] - I_{Y_{\Tilde{\phi}}}\fcal_\phi - K(\phi,\Tilde{\phi})
    \end{equation}
    which features a 2-cocycle $K$ which we will determine shortly.
    This is an on-shell bracket in that it is only antisymmetric on-shell. It is convenient to use the Noether charge as $Q_\phi$ when possible, in which case one speaks of the Noetherian split.\cite{freidelExtendedCornerSymmetry2021}
    The procedure of choosing the Noetherian split does not work here, however, due to the aforementioned schism between the Lagrangian and the Hamiltonian symmetry. We therefore need to find other criteria to argue for the \say{correct} charge bracket and/or split, and in particular the form of the corner charge. The simplest \say{charge bracket} one can come up with is just the double contraction of the symplectic form:
    \begin{equation}
    \{P_\phi,P_{\Tilde{\phi}}\}_{dc}    :=
    I_{Y_{\Tilde{\phi}}}I_{Y_\phi}\Omega = P^{(0)}_{\Phi(\phi,\Tilde{\phi})} 
    \end{equation}
    which reproduces \textit{only} the bulk part of the charge; Therefore, if we want to relate it to the above notion of an adapted bracket, the suitable split is to regard the corner charge as $Q_\phi=0$ and the flux as $\fcal_\phi=-\phi\cdot\alpha$.
    In particular, one would need to fix the cocycle $K$:
    \begin{equation}
        K(\phi,\Tilde{\phi})= \oint_{\partial\Sigma}(\phi\lor\Tilde{\phi})\cdot((F_\omega)_{(\beta)}-\Lambda\star\theta^2)
        .
    \end{equation}
    This split and cocylce choice then correctly implies that the charge bracket of vanishing corner charges vanishes.\\
    If instead we choose the BF-inspired corner charge, and use the same cocycle,
    \begin{equation}
        \begin{gathered}
            Q_\phi = -(\phi\lor\theta)_{(\beta)}\wedge\omega\\
            \fcal_\phi = \phi\cdot\gamma,
        \end{gathered}
    \end{equation}
    we get 
    \begin{equation}
        \{Q_\phi,Q_{\Tilde{\phi}} \}_c = Q_{\Phi(\phi,\Tilde{\phi})}
    \end{equation}
    which represents the gauge algebra exactly. Quite interestingly, the cocycle vanishes in the \say{ground state} of tetrad GR given by (Anti)deSitter spacetime (or flat Minkowski for $\Lambda=0$). In principle, these cocycles can have multiple origins such as non-covariance of the Lagrangian under the gauge transformation. The correct interpretation of it here is, however, up for debate.
    Yet another definition of an \say{effective bracket} is the antisymmetric flow of the charges
    \begin{equation}
        \{P_\phi,P_{\Tilde{\phi}} \}_{as} := \frac{1}{2}(Y_\phi[P_{\Tilde{\phi}}] - Y_{\Tilde{\phi}}[P_{\phi}]).
    \end{equation}
    which, if applied to the pure bulk charges, yields
    \begin{equation}
        \{P^{(0)}_\phi,P^{(0)}_{\Tilde{\phi}} \}_{as} = P^{(0)}_{\Phi(\phi,\Tilde{\phi})} -K(\phi,\Tilde{\phi})
    \end{equation}
    which admits an interpretation as an adapted charge bracket if one chooses a cocycle $K_{as}(\phi,\Tilde{\phi})= 2K(\phi,\Tilde{\phi})$ and $Q_\phi=0,\fcal_\phi=-\phi\cdot\alpha$.
    If instead applied to the BF-like charges, we get full closure of the algebra:
    \begin{equation}
    \{P_\phi,P_{\Tilde{\phi}} \}_{as} = P_{\Phi(\phi,\Tilde{\phi})}
    \end{equation}
    So we can interpret, for the BF-like charges, the antisymmetric flow bracket as the adapted charge bracket for the same charges. \\
    Overall we can see that due to the nonintegrability of the vector field $Y_\phi$, the usual subtleties regarding charge commutation relations arise that are already well-known from diffeomorphisms. In particular, a cocycle is needed in the on-shell charge bracket for consistency. It remains to be established which charge bracket is the most meaningful to use, a task which we leave to future work. \\
    Finally, if we apply the adapted charge bracket to the Lorentz/Translation commutation relations, we can fix the cocyle $K(\alpha,\phi)$ to be zero for consistency when $Q_\alpha=0$, and find for the BF-like charges
    \begin{equation}
        \{C_\alpha,P_\phi\}_c = P_{\alpha\cdot\phi} - \oint_{\partial\Sigma} (\phi\lor\theta)_{(\beta)}\wedge d\alpha
    \end{equation}
    which is again somewhat reminiscent of the corner algebra in 3D.\\

    In summary, we have presented the phase space canonical transformation $Y_\phi$ in tetrad gravity and its properties. We showed that it is generated by the Einstein constraint and that it generates the correct time evolution on the phase space, when combined with Lorentz transformations. Furthermore, it is not simply the pushforward of a transformation on the spacetime configurations. We also presented that the transformations and charges close with structure functions, and that care must be taken when dealing with nonintegrable contributions on corners.\\
    We now proceed with analysing the corner piece of the translation charges in detail.
    
 \section{Analysis of the corner charge and example}
    We thus have charges which form an analogue of a Poincare algebra, and which are manifestly Lorentz covariant. 
    Given that the charges and transformations we gave are intimately related to diffeomorphisms, an obvious question is how the corner piece relates to other ones found in the literature.
    There is an immediate way to do this for the charges corresponding to spatial translations: Adopting a 3+1 internal decomposition along the adapted internal normal $u_\Sigma$ to a slice $\Sigma$ (see \ref{App:Ashtekar} for notational details), we can expand the momentum $p^I$ into (here for now without the $\Bar{\omega}$ offset to highlight certain features) 
    \begin{equation}
        \begin{gathered}
                    p^I =-(\omega)^{IJ}_{(\beta)}\wedge\theta_J \\
                    =-\beta ((\Gamma_\parallel-\gamma K)\times \theta)^I - (\Gamma_\parallel+\beta K)^I \wedge (u\cdot\theta) +  u^I (\Gamma_\parallel+\beta K)_J\wedge\theta^J .
        \end{gathered}
    \end{equation}
    In this expression, the combinations of the spatial spin connection $\Gamma_\parallel$ and the extrinsic curvature $K$
    \begin{equation}
        A = \Gamma_\parallel-\gamma K \quad C=\Gamma_\parallel+\beta K
    \end{equation}
    are recogniseable as the Ashtekar-Barbero-Sen connection and its counterpart which is needed to reconstruct the full spin connection. What's noteworthy is that because these objects are connections, neither of the terms above can be Lorentz vectors. However, as soon as we subtract the $\Bar{\omega}$ offset, this is no longer an issue. We will only be interested in the value on spatial slices and particular on corners. On a given spatial slice (with $u_\Sigma^2=1$), $\Tilde{\theta} \stackrel{\Sigma}{=} \Tilde{e} $ reduces to the spatial triad, and we have
    \begin{equation}\label{spatialMomentum}
        \Tilde{p}^I 
            \stackrel{\Sigma}{=}-\beta ((\Tilde{\Gamma}_\parallel-\gamma \Tilde{K})\times \Tilde{e})^I +  u^I (\Tilde{\Gamma}_\parallel+\beta \Tilde{K})_J\wedge\Tilde{e}^J .
    \end{equation}
    Let us also further decompose the fields on a (closed, isolated) corner surface $S\subset\Sigma$, and decompose the triad there with respect to a spacelike 1-form $\mathbf{s}$ of the surface when included in $\Sigma$. Let $\hat{v}_i^I$ denote the inverse triad, and $\varsigma_I = \hat{v}_i^I s_i $ the internal adapted radial normal, then we decompose
    \begin{equation}
        \Tilde{e} \stackrel{S}{=} \Bar{e} - \varsigma \mathbf{s}
    \end{equation}
    into a zweibein $\Bar{e}$ and the radial piece. This implies through some calculations that the gravitational flux $E = \frac{1}{2}(\Tilde{e}\times \Tilde{e})$ splits similarly as
    \begin{align}
        \Tilde{E}^I &\stackrel{S}{=} \Bar{E}^I - \mathbf{s}\wedge E_r \\
        &= -\mathbf{w}\, \varsigma^I - \mathbf{s} (\varsigma\times \Bar{e})^I
    \end{align}
    where $\mathbf{w}$ is the area density 2-form of the zweibein $\Bar{e}$ on $S$, defined as
    \begin{equation}
        \mathbf{w} = |\text{det}(\Bar{e}^I_a)|d^2x.
    \end{equation}
    As $S$ is 2-dimensional, any 2-dimensional 2-form must be related to $\mathbf{w}$ through some prefactor. We can write this prefactor for $p^I$ by first splitting
    \begin{equation}
        A^I \stackrel{S}{=} A^I_J e^J \quad 
        C^I \stackrel{S}{=} C^I_J e^J,
    \end{equation}
    then it is simply
    \begin{equation}
        \Bar{p}^I  \stackrel{S}{=} 
        \left[
         \beta \varsigma_J (A_C^{\;\;C}\Tilde{\eta}^{IJ}-A^{JI}) + u^I \Tilde{\epsilon}^{UVW}\varsigma_U C_{VW}
        \right] \mathbf{w}.
    \end{equation}
    This is interesting as it suggests that the only nonzero components of $p$ on the corner are the timelike $u$ and radial $\varsigma$ ones, unless $A^{IJ}$ has off-diagonal elements as a 3x3 matrix. 
    It is worth comparing this to the corner Lorentz charge of tetrad gravity, evaluated in the same way, which reads
    \begin{equation}
        \theta^2_{(\beta)}\cdot\alpha \stackrel{S}{=} (u\lor\varsigma)\cdot(1-\beta\star)\alpha \, \mathbf{w}
    \end{equation}
    and which also only picks up the internal directions normal to the surface $S$.\\
    To get a better interpretation of this object, we go on-shell of the Gauss constraint, bringing us closer to metric gravity. In principle, this means that we replace $\Gamma_\parallel = \gamma[\tilde{e}]$ by the 3D Levi-civita spin connection
    \begin{equation}
        \gamma^I_i[e] = \frac{1}{2}\hat{v}^{Ij}\tilde{e}_{iK} (\partial_t \tilde{e}^K_j - \partial_j u^K)
    \end{equation}
    which one obtains via the Koszul formula\cite{freidelEdgeModesGravity2020} for the full 4D Levi-civita spin connection. On a slice $\Sigma$, it implies rather
    \begin{equation}
        \Tilde{d}_{\Gamma_\parallel}\tilde{e} = 0 = \Tilde{e}_I\wedge \Tilde{K}^I.
    \end{equation}
    By remembering that we should take differences of connections, we then can drop the $\Gamma_\parallel\times e$ and $K_I e^I$ terms in \ref{spatialMomentum}, and the expression simplifies to
    \begin{equation}
        \Bar{p}^I  \stackrel{S}{\cong} 
        \left[
         (K^{JI}- \Tilde{\eta}^{IJ} K_C^{\;\;C})\varsigma_J + u^I \Tilde{\epsilon}^{UVW}\varsigma_U \gamma[\Tilde{e}]_{VW}
        \right] \mathbf{w}
    \end{equation}
    and 
    \begin{equation}
        \gamma[\Tilde{e}]_{IJ} = \frac{1}{2}\hat{v}^{j}_I \Tilde{\eta}_{JK}
 (\partial_t \tilde{e}^K_j - \partial_j u^K).
    \end{equation}
    The object $(K^{JI}- \Tilde{\eta}^{IJ} K_C^{\;\;C})$ is simply the usual gravitational ADM momentum built from the extrinsic curvature tensor, so the term with $\varsigma$ is just the Brown-York charge\cite{freidelEdgeModesGravity2020,Brown:1992br}. In particular, for timelike translations, this term will not contribute and only the term involving $\gamma[\Tilde{e}]$ is going to be relevant. 
    
    A litmus test for the charges is whether they are nonzero for timelike translations: This is not the case for many other possible generalisations of the charges in 3D. The full \say{timelike} charge takes the form
    \begin{align}
        P_{f u} = &-\int_\Sigma f(u\lor\theta)_{(\beta)}{\wedge} F_{\hat{\omega}} 
        - \oint_{\partial\Sigma} 
        f (u\lor\theta)_{(\beta)}\wedge\hat{\omega}
        \\
        &= 
        -\int_\Sigma f \mathcal{H} 
        - \oint_{\partial\Sigma}  f\left[ \beta K_I \wedge \Tilde{e}^I - \Tilde{e}_I \wedge\Gamma^I_\parallel \right]
    \end{align}
    where the bulk piece is the Hamiltonian constraint and the boundary piece is the boost piece of the torsion constraint, \textit{plus a nonzero piece}:
    \begin{equation}
        \oint_{\partial\Sigma} f \Tilde{e}_I \wedge\Gamma^I_\parallel
    \end{equation}
    This involves only the 3D spin connection and the triad on the slice. \\
    For the timelike translation corner charge, we can use the identity
    \begin{equation}
        d_\Gamma E = dE - (\Gamma_\parallel\cdot e) \wedge e
    \end{equation}
    together with the Boost constraint $d_\Gamma e = 0$ to get an alternative, on-shell expression of the charge:
    \begin{equation}
        p_\perp \approx \rho_I E^I =  \text{div}_{e}(\hat{v}_I) E^I
    \end{equation}
    So, it is given by the scalar product of the gravitational flux with the internal vector $\rho$ whose components are given by the divergence of the inverse triad $\hat{v}_I$. This has the advantage of only depending on the triad itself and not on the connection. Then, on $S$ the timelike translation charge is
    \begin{equation}
        i_S^\ast p_\perp \approx -\text{div}_{e}(\hat{v}_I) \varsigma^I \mathbf{w}
    \end{equation}
    Note that even though this divergence form looks like one could extract a codimension 3 term from it using Stokes' theorem, the resulting puncture density would be proportional to
    \begin{equation}
        \varsigma^I \hat{v}^{rad}_I
    \end{equation}
    and so we should typically expect it to vanish. $rad$ here refers to the radial (spacetime) direction normal to a puncture, so within $S$.
    In principle, this does not immediately yield a geometric interpretation of the timelike charge, but is better than the off-shell form as it is expressed fully through tetrad data in a mostly uncomplicated form.\\
    \subsection{Schwarzschild corner charge density}
    In order to illustrate the content of this charge, we can calculate the full corner charge for example for the usual Schwarzschild spacetime. For this, we let
    \begin{equation}
        p^I = -\star\omega^{IJ}\wedge\theta_J
    \end{equation}
    be the \say{bare} ADM momentum aspect. Let 
    \begin{align}
        f(r) := \sqrt{1-\frac{2M}{r}}.
    \end{align}
    We then find in the diagonal Lorentz gauge, and in Schwarzschild coordinates:
    \begin{align}
        p^0 &= \frac{\cos{\theta}}{f(r)}dr\wedge d\phi - 2r\sin{\theta}f(r) d\theta\wedge d\phi \\
        p^1 &= -\cos{\theta}f(r) dt\wedge d\phi\\
        p^2 &=(1-\frac{3M}{r})\sin{\theta}dt\wedge d\phi\\
        p^3 &=-(1-\frac{3M}{r})dt\wedge d\theta
    \end{align}
    So we can see that this momentum is nonzero even in flat spacetime when $M=0$. We can then do the usual subtraction procedure and integrate it over different surfaces, most notably a sphere at spatial infinity. For this we first subtract the flat space contribution, and obtain 
    \begin{align}
        \oint_{S_{2,R}} p^I-p^I_{flat} 
        &= \delta^I_0\oint_{S_{2,R}} 2r \left[1-f(R)\right]\sin{\theta}d\theta\wedge d\phi  \\
        &= \delta^I_0 8\pi R\left[1-f(R)\right] \stackrel{I=0}{=} 8\pi M + 4\pi \frac{M^2}{R} + \mathcal{O}(\frac{1}{R}^2)
    \end{align}
    So for a sphere at infinity, the renormalised momentum aspect $p-p_{flat}$ points along the timelike direction (so describes an object at rest) and gives the black hole (ADM) mass up to a factor. Also, as expected, the subtracted $\theta-\phi$ component of $p^0$ agrees up to a factor of two with the Brown-York quasilocal surface energy density\cite{Brown:1992br}. It should be clear that this object depends on a reference connection to be evaluated, just like these other charges.\\
    In particular, on finite surfaces, the momentum can be integrated and vanishes on surfaces of constant $\theta,\phi$ and of constant $t,\phi$. It also satisfies the fact noted above that, if evaluated on any surface, the only nonzero components of the momentum are in the directions normal to the surface. E.g. for a $t-\phi$ plane, the only nonzero components are $p^1$ and $p^2 $ corresponding to directions $r$ and $\theta$. On the horizon, the renormalised momentum is \say{internal-angular} only:
    \begin{align}
        (p-p_{flat})^0 &=0 \\
        (p-p_{flat})^1 &= 0\\
        (p-p_{flat})^2 &=-\frac{3}{2}\sin{\theta}\; dt\wedge d\phi\\
        (p-p_{flat})^3 &=\frac{3}{2} dt\wedge d\theta
    \end{align}
    We can also overall tell that for this coulombic/nonradiative type of gravitational field, the charge has a peeling property in that up to a constant term picked up as the ADM mass, the remainder is subleading and scales as $\sim \frac{1}{r}$. Also, we point out that a very similar type of expression to our corner charge has been previously linked at asymptotic spacelike corners more generally to the ADM charges\cite{kabelQuantumReferenceFrames2023}.
    It would also be interesting to study the relation between the momentum charges described here and in the reference and the standard gravitational charges on \textit{null} surfaces, in particular to verify whether the charges here can be interpreted as the momenta of some gravitational radiation. We leave this task to future work.

 \subsection{Dynamical reference frames and hydrodynamics}\label{DynRefFr}
    Motivated by the form of the translation charges and form of their dependence on the connection $\hat{\omega}$, we will now show that they are better behaved on a certain phase space which is a slight extension of that of tetrad gravity. The theory can be given a spacetime Lagrangian as follows:
    \begin{equation}
        L = \theta^2_{(\beta)}\wedge F_{\Bar{\omega}+\Delta} - B\wedge F_{\Bar{\omega}}
    \end{equation}
    so consists of a flat BF Lagrangian for a connection $\Bar{\omega}$ and the Einstein-Cartan-Holst piece for an \textit{effective connection} \begin{equation}
        \omega:= \Bar{\omega}+\Delta
    \end{equation}
    where $\Delta$ is treated as an independent variable. By design, the equations for the tetrad and for $\Delta$ reproduce those of tetrad gravity, so that the theory contains gravity as a subsector. \\
    The part that we have extended by is a flat connection $\Bar{\omega}$ by which we \textit{dynamically split} the usual spin connection $\omega$. In this sense, we are providing, on a dynamical basis, a reference frame of a certain type. As per equivalence principle, this is locally always possible at a point. We will see that on this slightly extended phase space, which on-shell of the BF-constraints only differs from the tetrad gravity one on corners, we can find a parametrisation which transforms in a useful way, once we restrict back to the tetrad gravity phase space. This behaviour is also inherited by the charges. We believe that this theory is a useful starting point for discretisation of the charges we presented here.\footnote{It is noteworthy that the theory has certain similarities with teleparallel gravity\cite{aldrovandiTeleparallelGravityIntroduction2012}, where the usual affine connection of gravity is \textit{flattened} and curvature is equivalently stored in additional degrees of freedom. However, there is no immediate reason to interpret the variable $p$ as torsion here.}\\
    To begin the analysis, we have the presymplectic form of the theory:
    \begin{equation}
        \Tilde{\Omega}_\Sigma = \int_\Sigma\delta(\theta^2_{(\beta)}-B)\wedge\delta\Bar{\omega} + \delta\theta^2_{(\beta)}\wedge\delta\Delta
    \end{equation}
    The form has a kernel, which may be found by the full contraction with a vector field $X$:
    \begin{equation}
    \begin{aligned}
        0\stackrel{!}{=}-I_X\Omega_\Sigma = \int_\Sigma 
        &-\delta\theta_I \left[X[\Bar{\omega}]_{(\beta)}\wedge\theta+X[\Delta]_{(\beta)}\wedge\theta\right]^I \\
        &- \delta B\wedge X[\Bar{\omega}] + \left[X[B] -X(\theta^2)_{(\beta)} \right]\wedge\delta\Bar{\omega}\\
        &-X[\theta^2]_{(\beta)}\wedge\delta\Delta
    \end{aligned}
    \end{equation}
    The kernel then consists only of the $\Bar{\Delta}$-vector fields from before (see \ref{DeltaBar}), with
    \begin{equation}
        X[\Delta]_{(\beta)}\wedge\theta = 0.
    \end{equation}
    We can gauge fix this again by the structural constraint, applied to $\omega = \Bar{\omega}+\Delta$. This then allows us to do the following inversion: Given a vector-valued 2-form $p^I$ on $\Sigma$, there exists a \textit{unique} $\Delta$ in the $\Bar{\Delta}$-equivalence class such that the structural constraint holds for $\omega$ and
    \begin{equation}
        p = -\Delta_{(\beta)}\wedge\theta.
    \end{equation}
    This uniqueness allows, in particular, that we replace
    \begin{equation}
        -\delta\Delta_{(\beta)}\wedge\theta \mapsto \delta p
    \end{equation}
    in the (now nondegenerate) symplectic form, which includes our gauge fixing:
    \begin{equation}
        \Omega_\Sigma = \int_\Sigma\delta C\wedge\delta\Bar{\omega} + \delta\theta\delta p
    \end{equation}
    Here, we defined the coordinate
    \begin{equation}
        C := \theta^2_{(\beta)}-B.
    \end{equation}
    The important point is that $(\theta,p,\Bar{\omega},C)$ are independent coordinates for the phase space, whereas $\Delta,B$ are functions of them.\\
    The equations of motion of the theory are also easily found:
    \begin{equation}
        \mathcal{E}= -\delta B \wedge F_{\Bar{\omega}} - d_{\Bar{\omega}}( \theta^2_{(\beta)} -B)\wedge\delta\Bar{\omega}
        + \delta \theta \wedge G_\omega - d_\omega\theta^2_{(\beta)}\wedge\delta\Delta
    \end{equation}
    The first two fix the connection to be flat and $B$ to be the gravitational flux up to covariant $\Bar{\omega}$-derivatives, so no new local degrees of freedom are added. The last two are simply the usual tetrad gravity equations. From these, we can find the following four sets of charges, associated with their own gauge invariance each:
    \begin{equation}
        J_\alpha = \int_\Sigma C\wedge d_{\Bar{\omega}}\alpha
    \end{equation}
    \begin{equation}
        K_\mu = -\int_\Sigma \mu\wedge F_{\Bar{\omega}} - \oint_{\partial\Sigma}\mu\wedge\Bar{\omega}
    \end{equation}
    \begin{equation}
        C_\alpha = \int_\Sigma \theta^2\wedge d_{{\omega}}\alpha = \int_\Sigma \theta^2_{(\beta)}\wedge d_{\Bar{\omega}}\alpha - (\theta\wedge p)\cdot \alpha
    \end{equation}
    \begin{equation}
        P_\phi = -\int_\Sigma (\phi\lor\theta)_{(\beta)}\wedge F_{\omega} - \oint_{\partial\Sigma}\phi\cdot p
    \end{equation}
    The nontrivial transformations induced by these are, in order,
    \begin{equation}
        \{J_\alpha, \Bar{\omega} \} = d_{\Bar{\omega}}\alpha \quad \{J_\alpha,B \} = [B-\theta^2_{(\beta)},\alpha] \text{ or } \{J_\alpha,C \} = [C,\alpha]
    \end{equation}
    \begin{equation}
        \{K_\mu, B\}=-d_{\Bar{\omega}}\mu \text{ or } \{K_\mu, C\} = d_{\Bar{\omega}}\mu
    \end{equation}
    \begin{equation}
        \begin{gathered}
            \{C_\alpha,\theta\}=-\alpha\cdot\theta \quad \{C_\alpha, \Bar{\omega} \}= d_{\Bar{\omega}}\alpha \quad \{C_\alpha, p \}= -\alpha\cdot p
        \end{gathered}
    \end{equation}
    \begin{equation}
        \{P_\phi,\theta^2 \}= d_\omega (\phi\lor\theta) \quad \{P_\phi, p\}= - (F_\omega)_{(\beta)}\cdot\phi
    \end{equation}
    which in particular shows that by switching from $\Delta$ to $p$, we have made the new translation transformations explicit, rather than implicit. $\Delta$, as a matter of fact, still transforms as we saw it before:
    \begin{equation}
        \{C_\alpha,\Delta\} = [\Delta,\alpha] \quad \{P_\phi,\Delta \} = \mathbb{F}_\phi 
    \end{equation}
    with the \textit{same} $\mathbb{F}_\phi$ as we had before. In order to make the charges above the Hamiltonian generators of these transformations, we have to deal with the $P_\phi$ corner term as before. The relevant point is that we need to remove the \say{flux} term
    \begin{equation}
        \oint_{\partial\Sigma} (\phi\lor\delta\theta)_{(\beta)}\wedge\Delta
        -(\phi\lor\theta)_{(\beta)}\delta\Bar{\omega} .
    \end{equation}
    The first term is the same integrability condition we had before, which is unavoidable due to the open nature of the system. The second term, however, we can remove e.g. by imposing Dirichlet boundary conditions for our dynamical reference frame $\Bar{\omega}$. This means that we provide a partial reference frame on the corner and its time evolution, but inside the region $\Sigma$, it is determined dynamically from the flatness condition. \\
    The picture presented here has the following advantage: Previously, it was unclear how to interpret the corner term of the translation charges due to the connection being present, and furthermore it was not Lorentz invariant. Due to the specific (dynamical!) splitting here, we could write a Lorentz covariant corner piece in a way that does not jeopardise the nature of the phase space. Instead of fixing a generic connection $\omega_0$ as a reference point to define $p$, which would lead to $p$ not transforming as a Lorentz vector, we here have a dynamical split that does not restrict or add significantly to our phase space. Furthermore, this splitting of the connection into a flat piece and a momentum piece can be productive in studying different effects in gravity: Topological, BF-like effects can be encapsulated in the $\Bar{\omega}$-terms, while the local degrees of freedom are contained in the $\theta,p$ pair. In this way, we get the best of both the connection and the ADM momentum formulations of gravity.\\
    We also want to highlight that now there is another clear analogue of the translations of 3D gravity, generated by the \textit{commuting} charges
    \begin{equation}
        T_\phi = -\int_\Sigma p\wedge d_{\Bar{\omega}}\phi
    \end{equation}
    \begin{equation}
        \{T_\phi,\theta\} = d_{\Bar{\omega}}\phi
        ,\; \{T_\phi,B\} = p\lor\phi + (d_{\Bar{\omega}}\phi\wedge\theta)_{(\beta)}
    \end{equation}    
    In fact, under the Lorentz generators $C_\alpha$ (not to be confused with the \say{fake Lorentz} generators $J_\alpha$), these charges are vectors, so we have a true ultralocal Poincare algebra. What is now clear is that this Poincare algebra, while aesthetically simpler, is \textit{not} a gauge algebra: The bulk term of the charges $T_\phi$ amounts to $d_{\Bar{\omega}}p $, which is not the Einstein constraint, but only a linearised version of it. So these charges do not generate the gauge dynamics of gravity. Instead, one might see the \say{true} charges $P_\phi$ as a \say{collective} momentum charge since it involves $\omega$ as a function of $\Bar{\omega}$ and $p$. 
    This idea is strengthened further by the shape of the Einstein constraint when expanded into the new variables,
    \begin{equation}
        \begin{gathered}
            -G_\omega^I  = d_{\Bar{\omega}}p^I - \Delta_{(\beta)}^{IJ}\wedge d_{\Bar{\omega}}\theta_J -\frac{1}{2} \Delta^{I}_J\wedge (\Delta_{(\beta)})^J_K \theta^K - \frac{1}{2} (\Delta_{(\beta)})^I_J \wedge \Delta^{J}_K \theta^K  \\
            = d_\omega p^I - \Delta_{(\beta)}^{IJ}\wedge d_{{\omega}}\theta_J - \frac{1}{2}([\Delta,\Delta]_{(\beta)})^I_J \theta^J
        \end{gathered}
    \end{equation}
    which has explicitly the form of a conservation equation for $p$ up to a nonlinear term quadratic in $\Delta$ and a contribution which vanishes on-shell of the true Gauss constraint.\footnote{In fact, as is visible from the form we presented the second line in, the nonlinearity arises from the coefficient $\frac{1}{2}$ in the definition of the curvature. In other words, if the curvature was $d\omega + [\omega,\omega]$, the Einstein equations would be exactly a conservation equation (still with implicit nonlinearities). }
    Weighing this by the gauge parameter, the above turns into
    \begin{equation}
        0\approx -\phi_I G_\omega^I = d(\phi_I p^I) - Y_\phi[\theta]_I\wedge p^I - \frac{1}{2}([\Delta,\Delta]_{(\beta)})\wedge(\phi\lor\theta)
    \end{equation}
    which demonstrates explicitly that the constraint is a nonconservation/flux-balance law for the charge $\phi\cdot p$, which should hold on codimension 1 surfaces. This is reminiscent of ideas from the fluid/gravity correspondence, in particular the membrane paradigm and AdS/CFT\cite{damourSurfaceEffectsBlackHole1982,thorneBlackHolesMembrane1986,Damour:2008ji,Hubeny:2011hd,Kolekar:2011gw,Janik:2005zt,Bhattacharyya:2007vjd}.
    From this point of view, the charges $T_\phi,C_\alpha$ provide a simple \say{microscopic} description of the system, whereas $P_\phi,C_\alpha$ describe its \say{effective macroscopics}. The relation between the charges $P_\phi$ and $T_\phi$ is intriguing and requires further study, which we postpone to future work.\\

    In summary, we have analysed the content of the corner translation charge $\phi_I p^I$ on a 2-dimensional corner and related it to the Brown-York charge. Importantly, there is a difference for timelike translations. In particular, it is nonvanishing for timelike translations. We then evaluated the charge on the familiar Schwarzschild geometry and demonstrated the typical peeling property, yielding the ADM mass at spacelike infinity. Building on the fact that the charge requires an offset to be meaningful, we extended tetrad gravity by a flat connection as a dynamical reference frame to use as an offset, and presented the adjusted charges and transformations. We also noted, finally, that there seems to be a simpler version of the translation generators in a specific regime.\\
    We now move forward with an extended discussion.

\section{Discussion}
    Various approaches to quantisation and discretisation of gravity have used 3D tetrad gravity as a benchmark for many technical and conceptual issues of the field. Of course, due to the simplicity of the 3D case, many properties of it are commonly held to not carry over to the more physically interesting 4D case. Chief among them is the presence of nontrivial \textit{local} (per-point) degrees of freedom given by gravitons. The 4D case becomes the most similar to 3D when global flatness is imposed\cite{Asante:2018kfo}, in which case the theory is exactly solvable in-vacuo. The only degrees of freedom in this sector are edge modes: Lorentz group elements $g$ from the connection and $\rbb^{1,3}$ elements $\phi$ from the tetrad. This demonstrates elementarily that there is an analogue of the triad shifts from 3D also in the 4D case. What is remarkable is that there is indeed an analogue of these transformations on the entire phase space of tetrad gravity in 4D, rather than just in special cases like the flat sector. In this paper, we have presented a careful examination of these transformations in order to make them more accessible to further studies in classical and quantum gravity, agnostic of approach.\\
    In this extended discussion, we first make some general remarks before illustrating some implications for the \textit{edge modes} of tetrad gravity, understood as dynamical reference frames living on timelike screens that are evolutions of corners. These edge modes generically carry representations of the corner symmetry group, which is what we focus on. After that, we discuss the inclusion of matter, in particular the previously untreated case of Dirac spinors, for whose curved spacetime treatment the connection+tetrad formalism is indispensible. We then close by mentioning possible avenues of extension for this work.\\
    
    Curiously enough, the charges of tetrad gravity derived in the way we presented here are still \textit{quadratic} in the sense of being constructed from two copies of $\theta$ and $\omega$, unlike in the 3D case where the corner charges are linear in the fields. We could therefore imagine also a set of charges which has corner value weighted by 1-forms,
    \begin{equation}
    \oint_{\partial\Sigma}\theta_I\wedge\rho^I
        \qquad
        \oint_{\partial\Sigma} \omega\wedge\mu.
    \end{equation}
    However, the latter is actually the corner charge of BF theory's flatness constraints, and therefore can only be differentiable in certain sectors; and the former can be understood as a dressed version of the Lorentz charges which use $\theta^2_{(\beta)}$. Therefore we needn't worry about the lack of linear $\theta$-charges, while the lack of linear $\omega$-charges is quite possibly a \textit{feature} of tetrad gravity.\\
    We would like to highlight the technical complexity necessary to study these new \say{nonlinear translations}: If one were coming from a purely Lagrangian point of view, one would have falsely assumed that the symmetry was no good due to its lack of a corner charge, its complexity, and the failure of projectability onto the phase space of any Cauchy slice. On the other hand, the phase space situation is similarly involved and carries integrability issues with it that would have frustrated many that would like to use them. Even having said that, the resulting transformations and charges are complicated and require various implicit definitions. It is therefore, in our view, not surprising that they have so far received only limited attention.\\
    Also, it is worth stressing that the status of the translations as \say{symmetries} of tetrad GR is subtle: Due to the mismatch of Lagrangian and phase space transformations, the canonical translations are certainly not \say{symmetries of the Lagrangian}. However, they are \textit{used} in writing the \textit{Hamiltonian} of the theory as
    \begin{equation}
        H = D_{\hat{n}} = C_{i_{\hat{n}}\omega} + P_{i_{\hat{n}}\theta}
    \end{equation}
    which suggests that they may have the status of a \say{symmetry of the Hamiltonian} instead. Care must be taken to distinguish the two notions here. In particular, the charges are not conserved generically due to their nonintegrable nature. \\
    However, we stress their implicit usefulness: Control over the 4D translations, as presented here, means control over the dynamics of general relativity. They are in several ways analogous to the Bergmann-Komar algebra of the ADM formalism\cite{thiemannModernCanonicalQuantum,bergmannPhaseSpaceFormulation1980} in that their action is complicated, but specifically tied to a certain form of dynamics.\\

    \subsection{Implications for edge modes}
    There are a few immediate implications from the existence of the above charges. The first is that the full corner symmetry modes of internal gauge transformations of tetrad gravity is on-shell some variation of
    \begin{equation}
        so(1;3)^{\partial\Sigma}\ltimes H
    \end{equation}
    where $H$ is the set $(\rbb^{1,3})^{\partial\Sigma}$ equipped with the stucture functions we found. Naively, this would lead to an overall corner potential for the pure-gauge corner degrees of freedom of the form
    \begin{equation}
        \oint_{\partial\Sigma} p_I \chi(\phi)^I(\hat{\omega},\theta) + \theta^2_{(\beta)}\cdot \chi(g)
    \end{equation}
    where $\chi(g),\chi(\phi)$ are Maurer-Cartan forms on the Lie group corresponding to the corner symmetries, parametrised by a Lorentz group element  per point $g$ and an element $\phi$ of $H$. $\chi(g)$ does not depend on the other fields, incorporating the field independence of the structure constants of the Lorentz symmetry. However, $\chi(\phi)$ is allowed to depend on the fields as to reproduce the field-dependent structure functions. The densities $\theta^2_{(\beta)}$ and $p = -\hat{\omega}_{(\beta)}\wedge\theta$ are then the pendents of left-invariant vector fields on the infinite-dimensional corner symmetry group.\\
    We can make this more precise by doing a split of the symplectic potential a la Gomes et Riello:\cite{gomesUnifiedGeometricFramework2019,gomesObserverGhostFieldspace2017}
    We split any differential (as a 1-form on field space) into  \textit{horizontal and vertical parts}, the former of which has only non-gauge variations in it. This is possible by introduction of additional structure in the form of a field space connection $\chi$  for the gauge group $\mathcal{G}$ (valued in $Lie(\mathcal{G})$ ) satisfying for the gauge vector fields $X_\alpha$ 
    \begin{equation}
        I_{X_\alpha}\chi = -\alpha, \quad L_{X_\alpha}\chi = -\delta_\chi \alpha := -(\delta\alpha + [\chi,\alpha]_\mathcal{G}) \quad \forall \alpha \in Lie(\mathcal{G}).
    \end{equation}
    These properties allow to define the horizontal derivative
    \begin{equation}
        \delta_\chi V := \delta V + L_{X_\chi}V \quad \implies I_{X_\alpha}\delta_\chi V = 0
    \end{equation}
    which only contains the non-gauge variations. This allows for a neat split of the symplectic potential
    \begin{equation}
        \theta^2_{(\beta)}\wedge( \delta_\chi \hat{\omega} - L_{X_\chi}\hat{\omega})
    \end{equation}
    which for our case has a gauge group with parameters for Lorentz and translation transformations, and we can split $\chi = (\zeta,\eta) $ into Lorentz and translation parts. Then the gauge piece is (here for simplicity with $\Lambda=0$)
    \begin{equation}
        L_{X_\chi}\hat{\omega} = d_\omega \zeta + \mathbb{F}_\eta  
    \end{equation}
    which yields the pure-gauge potential
    \begin{equation}
        \theta_g = (\eta\lor\theta)_{(\beta)}\wedge F_\omega - \theta^2_{(\beta)}\wedge d_\omega\zeta.
    \end{equation}
    Any other polarisation of the potential differs by a differential $\delta f$ for some function $f$ on field space, so that the full general gauge potential is 
    \begin{equation}
        \theta_g - L_{X_\chi}f.
    \end{equation}
    An interesting choice of this might be 
    \begin{equation}
        f= -\theta^2_{(\beta)}\wedge (\hat{\omega}-\omega_0) \qquad\delta\omega_0 = 0
    \end{equation}
    which gives the gauge potential
    \begin{equation}
        \theta_g = [\theta^2_{(\beta)},\zeta]\wedge(\hat{\omega}-\omega_0) + d_\omega(\eta\lor\theta)_{(\beta)}\wedge(\hat{\omega}-\omega_0).
    \end{equation}
    While the form for $f=0$ has a corner piece $-\theta^2_{(\beta)}\cdot \zeta$ for Lorentz transformations, this other form suggests a corner term for translations of the form
    \begin{equation}
        (\eta\lor\theta)_{(\beta)}\wedge(\hat{\omega}-\omega_0) = \eta_I p^I
    \end{equation}
    where the momentum aspect $p$ is here with respect to $\omega_0$. This is precisely as advertised in the beginning - the momentum $p$ becomes the corner generator for translations, while it is $\theta^2_{(\beta)}$ for Lorentz transformations.
    With this split (which has in general provided additional data to the theory in the form of a gauge fixing procedure), we isolate the non-gauge data in the horizontal piece and pick out the edge mode data in an easily readable fashion. We can also apply it to the symplectic form straightforwardly and pick out the corner terms, which yields the 1-parameter family\footnote{The 1-parameter freedom is due to the ambiguity in integrating a term of the form $d\alpha\wedge d\beta$ into $d(\alpha\wedge d\beta)$ or $d( d\alpha \wedge \beta)$.}
    \begin{equation}
        \begin{gathered}
            \omega_c = -\delta_\chi \theta^2_{(\beta)}\cdot\zeta + \theta^2_{(\beta)}\cdot \frac{1}{2}[\zeta,\zeta] - (\eta\lor\theta)_{(\beta)}\cdot\delta_\chi \hat{\omega} -(\eta\lor\eta)_{(\beta)}\cdot F_\omega \\
        +\lambda (\eta\lor\theta)_{(\beta)}\wedge d_\omega\zeta
        +(1-\lambda) d_\omega(\eta\lor\theta)_{(\beta)}\cdot \zeta
        \end{gathered}
    \end{equation}
    This expression contains, in principle, all there is to know about the edge modes: The expression for their charges, their commutation relations, and their couplings to the non-gauge degrees of freedom. In particular, we know that gravity in 4D has locally propagating degrees of freedom, which will couple kinematically via the $\delta_\chi \theta,\delta_\chi \omega$ terms here. In particular, the commutation relations of the supposed charges are 
    \begin{equation}
        \begin{aligned}
        I_{X_{(\Tilde{\alpha},\Tilde{\phi})}}I_{X_{(\alpha,\phi)}}\omega_c
            &= \theta^2_{(\beta)}\cdot[\alpha,\tilde{\alpha}] - (\phi\lor\Tilde{\phi})_{(\beta)}\cdot F_\omega \\
            &+ \lambda\left[
            (\phi\lor\theta)_{(\beta)}\wedge d_\omega\tilde{\alpha}
            -
            (\tilde{\phi}\lor\theta)_{(\beta)}\wedge d_\omega{\alpha}
            \right]\\
            &+ (1-\lambda) \left[
            d_\omega(\phi\lor\theta)_{(\beta)}\cdot\tilde{\alpha}
            -
            d_\omega(\tilde{\phi}\lor\theta)_{(\beta)}\cdot {\alpha}
            \right]\\
            &\hat{=} \{Q_{(\alpha,\phi)}, Q_{(\tilde{\alpha},\tilde{\phi})} \}
        \end{aligned}
    \end{equation}
    which shows clearly that on the corner, the bulk translations should not commute, but give a corner term of the form $(\phi\lor\Tilde{\phi})_{(\beta)}\cdot F_\omega$. Also, we can directly get hints for the form of the corner charges, as
    \begin{equation}
        I_{X_{(\alpha,\phi)}}\omega_c = \delta(\theta^2_{(\beta)}\cdot\alpha) + (\phi\lor\theta)_{(\beta)}\wedge\delta\hat{\omega}
        - d( (\phi\lor\theta)_{(\beta)}\cdot\zeta)
    \end{equation}
    If, once again, we manage to handle the issue of the $\theta\delta\omega$-term, the first two terms suggest once again
    \begin{equation}
        Q_{(\alpha,\phi)}=\theta^2_{(\beta)}\cdot\alpha + p_I \phi^I 
    \end{equation}
    and the codimension 3 piece drops out in integration. We can therefore be reasonably confident in saying that this is an appropriate general form for the corner charges of tetrad gravity in order to reproduce the correct commutation relations. \\
    We stress here that this pertains only to the part of phase space regarding the gauge edge modes, and that there are \textit{non-gauge corner degrees of freedom} which have nontrivial Poisson brackets with the edge modes; case in point, we can see that non-gauge variations of $\theta^2$ can have a nontrivial generator on the corner that possibly involves the edge mode $\zeta$.
    
    As a second point, we can use this information to ask for boundary conditions that allow the edge modes to be unrestricted. This requires noncovariant conditions in the form of
    \begin{equation}
        \begin{gathered}
            \hat{\omega}_t = F[\omega_a,\theta_a]\\
            \theta_t = G[\omega_a,\theta_a]
        \end{gathered}
    \end{equation}
    where the hat, so the gauge fixing, is important. These boundary conditions are in service of allowing $\theta_{a},\omega_a$ for indices $a=\theta,\varphi$ tangential to the surface $\partial\Sigma$ to be free, and similarly for $p_{\theta,\varphi}$. With these degrees of freedom unrestricted, so is $(\theta^2)_{\theta,\varphi}$, and so the corner charges have no restrictions on their densities.\\
    Furthermore, we can then also study the set of edge mode parameters $\phi,\alpha$ on the timelike screen $\partial\Sigma\times \rbb$ such that the corner charges $C_\alpha,P_\phi$ are conserved. This will yield PDEs for the edge mode fields that depend on the exact boundary conditions $F,G$ and are always of the form
    \begin{equation}
        q_{\theta\varphi}\cdot \partial_t  f = L[f;\theta_a,\omega_a]
    \end{equation}
    where $q$ denotes the corner charge density and $L$ some spatial pseudodifferential operator. The easiest option is $F=G=0$ in which case we can expect $L=0$, so all time-constant edge modes can be used as parameters to get conserved charges (assuming integrability).\\
    The appearance of structure functions signals that perhaps a purely \say{abstract} treatment of edge modes and corner symmetries is insufficient for complicated theories like gravity. Either, the phase space of corners must include geometric, \textit{non-gauge} variables $\theta_a,\omega_a$ as well as edge modes, or the structure functions must emerge at an effective level, for instance after a coarse graining procedure. The former could for example be realised on the on-shell level by having a corner metric be part of the phase space, and the non-gauge parts of $\theta,\omega$ as functions of it. 

        \subsection{Coupling to spinning matter}
    We would like to make a few remarks about the coupling to spinoral and other torsion-sourcing matter and its influence on the symmetry presented here. It is known from Montesinos et al.\cite{montesinosReformulationSymmetriesFirstorder2017} that the translation symmetry is highly sensitive to the addition of matter, particularly nonminimally coupled types. Additionally, due to the field-dependence of the structure functions, once nontrivial torsion is present on-shell, we have in principle different structure functions from the uncoupled case. For the purpose of illustrating this, we introduce the following Lagrangian for Dirac spinor fields $\Psi,\Bar{\Psi}$:
    \begin{equation}
        L_D = (\star\theta^3)_I \wedge\Bar{\Psi} i \gamma^I d_\omega \Psi - m \Bar{\Psi}\Psi \theta^2\wedge\star\theta^2
    \end{equation}
    The variations with respect to the spinors lead to 4-form equations of motion which do not entail new gauge invariances. However, the Einstein constraint receives a dynamical source:
    \begin{equation}
        \begin{gathered}
            G_\omega - \Lambda\star\theta^3 + N \approx 0\\
            N^I := (\star\theta^2)^{IJ}\wedge\Bar{\Psi} i \gamma_J d_\omega \Psi - m \Bar{\Psi}\Psi \star \theta^3
        \end{gathered}
    \end{equation}
    In particular, a condensate of massive fermions can mimick a cosmological constant. Similarly, the Gauss constraint receives a spin source:
    \begin{equation}
    \begin{gathered}
        S -d_\omega\theta^2_{(\beta)} \approx 0\\
        S^{IJ} := (\star\theta^3)_K \wedge\Bar{\Psi} i \gamma^K  \Sigma^{IJ} \Psi
    \end{gathered}
    \end{equation}
    where $\Sigma^{IJ}$ are the spin generators of the Clifford algebra.
    By modifying the constraint functions appropriately, we can get the appropriate modification of the translation and Lorentz symmetries. However first, we need to take into account, again, the possible presence of a nontrivial kernel in the symplectic form. We find by the same analysis that the same kernel for $X(\omega)$ is present, and by making use of Clifford relations, that there is no nontrivial kernel for the spinors.\footnote{The corresponding analysis involves using the fact that $u_\Sigma$ has nonzero square, making $u_\Sigma\cdot\gamma$ invertible. If it were not, there would be a nontrivial kernel for the spinors. If so, this also affects $X(\omega)$ to contain a certain term in order to stay in the kernel.} We can therefore use the same structural constraint as before. The Lorentz generator takes the form
    \begin{equation}
        C_\alpha = \int_\Sigma \theta^2_{(\beta)}\wedge d_\omega\alpha + \star\theta^3_I \Bar{\Psi}i\gamma^I (\alpha\cdot\Sigma)\Psi
    \end{equation}
    and generates the expected
    \begin{equation}
        X_\alpha[\Psi] = (\alpha\cdot\Sigma)\Psi
        \quad
        X_\alpha[\Bar{\Psi}] = 
        \Bar{\Psi}(\Sigma\cdot\alpha),
    \end{equation}
    the rest being unaffected. For the translations, we now have
    \begin{equation}
        P_\phi = \int_\Sigma (\phi\lor\theta)_{(\beta)}\wedge F_{\hat{\omega}} - \Lambda\phi\cdot\star\theta^3 + \phi\cdot N - \oint_{\partial\Sigma
        }(\phi\lor\theta)_{(\beta)}\wedge\hat{\omega}
    \end{equation}
    and the resulting transformations are quite different:
    \begin{equation}
        \begin{gathered}
            Y_\phi[\theta^2]_{(\beta)} = d_\omega(\phi\lor\theta)_{(\beta)} + \Bar{\Psi}\left((\star\theta^2)\cdot(\phi\lor \gamma)\right) i\Sigma\Psi\\
            Y_\phi[\omega]_{(\beta)}^{IJ}\wedge\theta_J = - \left[(F_\omega)_{(\beta)} - \Lambda(\star\theta^2) + \star((\Bar{\Psi}i\gamma d_\omega\Psi)\wedge\theta) - m\Bar{\Psi}\Psi \star\theta^2 \right]^{IJ}\phi_J\\
            Y_\phi[\Psi] = \phi^I \nabla_I \Psi - m (u_\Sigma\cdot\phi)(u_\Sigma\cdot\gamma)\Psi
        \end{gathered}
    \end{equation}
    where we expand the covariant derivative
    \begin{equation}
        d_\omega\Psi = \nabla_I \Psi \theta^I
    \end{equation}
    and omit the transformation for $\Bar{\Psi}$. Additionally, in the first line, the free indices lie on $\Sigma$.
    Multiple things are easily noticable here: The general form of the connection transformation law stays the same as a
    \begin{equation}
        Y_\phi[\omega]_{(\beta)}^{IJ}\wedge\theta_J = - M^{IJ}\phi_J
    \end{equation}
    with Lie algebra valued 2-form $M$. The transformation law for the gravitational flux $\theta^2$ changes drastically, mixing geometry and spinorial matter, but curiously in the form
    \begin{equation}
        Y_\phi[\theta^2] \sim d_\omega(\phi\lor\theta) + \theta^2\cdot \mathcal{M}
    \end{equation}
    with some 0-form $\mathcal{M}$ with 2 pairs of Lie algebra indices. Therefore, the spinorial contribution acts only internally. Furthermore, the transformation law for the spinor includes the \say{diffeomorphism} term $\phi^I\nabla_I$ which has already been noted by Montesinos et al. in the case of a minimally coupled scalar field, but also has unavoidable \say{non-covariant} contributions coming from the mass term.\\
    In principle, then, we can also expect the structure functions of the theory to change upon inclusion of matter. We do not attempt to derive this here. Instead, we highlight that this is a double-edged sword: The symmetry presented here is quite involved and sensitive to the details of the theory, but this can be seen as an improvement over diffeomorphisms or Lorentz transformations: These \textit{kinematical symmetries} are \say{too generic} to account for anything that is particular to GR as a dynamical theory. Therefore, imposing diffeomorphism and Lorentz invariance for a quantum theory, even with the \say{correct} set of kinematical data, does \textit{not} single out GR as its corresponding dynamics. Yet, the symmetry presented here, if implemented, would precisely do that.\\
    We stress, to avoid confusion, that this is a \textit{generic} phenomenon in theories whose evolution is 'pure-gauge', so where diffeomorphisms are given as field-dependent gauge transformations. Such theories include all standard topological theories like BF, Chern-Simons, but also all diffeomorphism-invariant theories of metrics like Lanczos-Lovelock gravities. In contrast, gauge theories like Maxwell theory or Yang-Mills have additional, \textit{non-gauge} evolution pieces (known as covariant diffeomorphisms) which do not fall into this class. In the case of 'full gauge' theories like BF and gravity, then, these covariant diffeomorphisms can be understood as \textit{additional} gauge symmetries (on top of some generic kinematical symmetry).\\
    Consider, as an example, BF theory with cosmological constant $\Lambda$, where the corresponding role is played by the Kalb-Ramond translations
    \begin{equation}
        Y_\mu[B]= d_\omega \mu \quad Y_\mu[\omega]= \Lambda \mu.
    \end{equation}
    Any diffeomorphism can be written on-shell as a combination of the internal gauge group (the pendent to Lorentz transformations) and these translations. But in contrast to the Lorentz transformations, these symmetries depend sensitively on the \textit{bulk term including the cosmological constant}. As such, the concrete form of the symmetry vector field actually singles out a given dynamics for the theory, unlike the Lorentz symmetry. \\
    The main difference between this simple case and tetrad gravity is, then, that the gauge group of gravity can not be studied abstractly, but must be understood as acting on a concrete given phase space of tetrads and connections. This simply reflects that unlike in BF theory, where all local dynamics is essentially trivial, tetrad gravity has locally propagating degrees of freedom. 
    
    \subsection{Outlook}    
    We would like to close by mentioning possible avenues of extending this work. First, more care is needed to study the behaviour of the nonintegrable (open boundary condition) case. Also, more light on the relations to existing charges such as diffeomorphism or the $SL(2;\rbb)$ groups\cite{freidelBubbleNetworksFramed2019,freidelEdgeModesGravity2020a} found on corners in various formulations of gravity is needed. Furthermore, the dynamics of the edge modes of tetrad gravity deserves its own study, which would likely bring with it much improved understanding of the impact of different boundary conditions on the integrability of charges. \\
    As a second larger direction, we propose to study various discretisations, truncations and quantisations of the transformations and charges presented here. One particular advantage of the new translations is that their parameter is not, like for diffeomorphisms, given by a vector field. While vector fields are difficult to discretise in an abstract fashion, the 0-form gauge parameters of the translations can simply be truncated to their values on vertices of whatever type of discretisation used. Being given gauge generators $C_\alpha, P_\phi$ per discretisation cell, one could then reconstruct analogues of diffeomorphism generators as higher order charges.\\
    On the other hand, by employing the right kinds of truncations of the phase space on small enough regions of space(time), the complications arising in the translations may become more tractable. This does not need to happen on the level of charges per se, but can also be about the full phase space of a small region.
    Additionally, in a \say{local holography} type approach, where small regions of space are represented by corner data and complicated macroscopic configurations like typical gravity arise as a coarse-grained limit of these data\cite{freidelGravitationalEnergyLocal2015,freidelEdgeModesGravity2021}, the bulk vanishing of these charges can be taken as \textit{imposing} the Einstein equations in the bulk of the elementary, small regions. This has been studied in the setting of asymptotic future null infinity\cite{freidelWeylBMSGroup2021a}, and possibly can be extended to finite regions, for example in a covariant way using the charges described here.\\
    In the same direction, we think that the formulation outlined in section \ref{DynRefFr} is particularly amenable to discretisation due to the simple form of its symplectic form - it is part BF theory, part relativistic particle. We expect that a lattice/triangulation discretisation, either in a direct form or by restriction to distributional configurations \`a la Regge calculus, will lead to a mixture of spinfoam and Regge geometry-like data which may be combined in various useful ways. \\
    In particular, we conjecture that discretisation of the transformation $Y_\phi$ in these variables is feasible and therefore allows determining a discretisation of the charge $P_\phi$, which includes the Hamiltonian and spatial diffeomorphism constraints in a covariant fashion. The data may therefore be well suited for a discrete, canonical analysis of gravity, but also for establishing relations to covariant path integrals of the Regge and spin foam types. As these discrete gravity approaches rely directly on them, it is important to identify the degrees of freedom of finite regions needed to properly model their dynamics as well as their glueing. In particular, typical spin foam models (such as the BC and EPRL-FK models) use kinematical Lorentz-BF type data, but the results here suggest the extension of these data to include edge vectors built from integrated tetrads. \\
    Finally, these discretisations may lend temselves well to continuations of existing work in the framework of Group Field Theories\cite{Oriti:2007qd}, which, motivated ab initio from independent arguments, also present a picture of spacetime as an emergent, hydrodynamical regime of a system of microscopic constituents\cite{Oriti:2010hg,Oriti:2024qav}. This links together quite naturally with the interpretation of the Einstein equations as hydrodynamics outlined before, and has been further validated in the cosmological subsector\cite{Oriti:2024qav}, where from \textit{yet further} arguments\cite{Lidsey:2013osa}, one expects such a hydrodynamical interpretation.\\
    Therefore, we believe that a rich conflux of methods and arguments coming from canonical quantisation, holographic and fluid/gravity correspondences, lattice gravity path integrals and Group Field Theories is possible using the results we presented in this work.
    
    \section*{Acknowledgements}
    SL would like to thank G.Neri for proofreading the document. DO acknowledges financial support from the ATRAE programme of the Spanish Government, through the grant PR28/23 ATR2023-145735.
    
    \appendix
    \section{Conventions}\label{App:Conventions}
    We collect here a few important conventions on the notation.\\
    Starting from section 2, we work with fields over a 4-dimensional, pseudo-Riemannian base $M$, possibly with  boundary. We equip this with a principal bundle $P\rightarrow M$ with structure group either the (identity component of the) Lorentz group $SO(1,3)$ or its spin cover $SL(2;\cbb)$, understood as the spin lift of the frame bundle $Fr(TM)$. We then associate a 'fake' tangent bundle $V$ to it via the usual 4D representation $\rho$ of these groups, $V:= P \times_\rho \rbb^{1,3}$.
    We can then choose a local frame $\{b_I\}_{0\leq I\leq 3}  \in \Gamma(V)$ for this fake tangent bundle and express $V$-valued sections locally through their components with indices $I = 0,1,2,3$. We equip $V$ with a nondynamical bundle metric $\eta$ and assume it is oriented.  \\
    We will also denote the Lie algebra of the structure group by $\gfrak$ and use several times that there is an isomorphism
    \begin{equation}
        \gfrak=\mathfrak{so}(1,3) \cong \Lambda^2(\rbb^{1,3})
    \end{equation}
    between antisymmetric bivectors and the Lie algebra. In this way, we can dispense with the Lie algebra and mostly work with wedge products in the fibres of $V$. \\
    We distinguish, when one of them is a 0-form in spacetime, between the spacetime wedge product of differential forms, $\wedge$, and the internal wedge product of sections of $V$, $\lor$. E.g. 
    \begin{equation}
        \phi\lor \theta \text{  vs.  } d_\omega\phi\wedge \theta
    \end{equation}
    has only an internal wedge product on the left, but \textit{both} internal and spacetime wedge product on the right.\\
    We heavily use a standard inner product on the Lie algebra $\gfrak$. When writing elements out in components with respect to a basis $M_{IJ} = b_I\lor b_J$, 
    \begin{equation}
        X\cdot Y = \frac{1}{2} X^{IJ}Y_{IJ} 
    \end{equation}
    where indices are raised and lowered with respect to the Lie algebra metric
    \begin{equation}
        \eta_{IJ,KL} = \eta_{IK}\eta_{JL}-\eta_{IL}\eta_{JK}.
    \end{equation}
    As bivector-valued objects $A,B$ are naturally identified with Lie algebra valued objects, we can apply this inner product whenever two bivectors appear in the same expression. In particular, when writing wedge products in integrals, we most often want scalar integrands, and adopt the convention 
    \begin{equation}
        A\wedge B := \frac{1}{2} A^{IJ}\wedge B_{IJ}
    \end{equation}
    which includes both wedge product and inner product.\\
    We also note the common use of the internal Hodge dual (with suppressed volume form on $V$):
    \begin{equation}
        (\star M)_{IJ} = \frac{1}{2}\epsilon_{IJKL}M^{KL} 
    \end{equation}
    and similarly for internal multivectors of other degree. Quite relevant is also the Lie-algebra commutator (just the matrix commutator)
    \begin{equation}
        [ A,B ]^{IJ}:= A^{I}_K\wedge B^{KJ} - (-1)^{|A|\, |B|} B^I_K\wedge A^{KJ}
    \end{equation}
    which satisfies\cite{cattaneoReducedPhaseSpace2019}
    \begin{equation}
        \star [ A,B ] = [ \star A,B ] = [ A,\star B ].
    \end{equation}
    We use for Lie algebra actions on vectors the simple matrix-vector product notation
    \begin{equation}
        (\alpha\cdot v)^I = \alpha^I_J v^J
    \end{equation}
    which, in context, should not be confused with the Lie algebra inner product, and for covariant derivatives of vectors
    \begin{equation}
        d_\omega v^I = d v^I + \omega^I_J \wedge v^J.
    \end{equation}
    For bivector/Lie algebra valued objects, in turn,
    \begin{equation}
        d_\omega A = d A + [\omega,A].
    \end{equation}

    \section{Symplectic vector fields }\label{SympVFDetail}
    
    Overall, with the gauge fixing from the structural constraint in place, the phase space of tetrad gravity is now 24-dimensional and has a nondegenerate symplectic form. We show here the general form of symplectic vector fields on this phase space. For this, first note the contraction
    \begin{equation}\label{Contraction}
        -I_X \Omega = \int_\Sigma X[\theta]_I  \wedge \delta\hat{\omega}^{IJ}_{(\beta)}\wedge\theta_J - \delta\theta_I \wedge X[\hat{\omega}]^{IJ}_{(\beta)}\wedge\theta_J.
    \end{equation}
    $X$ will then be symplectic if $\delta(I_X\Omega)=0$. Let us introduce the following shorthands:
    \begin{equation}
        \begin{gathered}
            A_{IJ} = \frac{\delta X[\theta]_I}{\delta\theta^J} \qquad 
            B_{IRD}= \frac{\delta X[\theta]_I}{\delta \hat{\omega}^{RD}}\\
            U_{RSK}= \frac{\delta X[\hat{\omega}]_{RS}}{\delta\theta^K} \qquad
            V_{RSAB}= \frac{\delta X[\hat{\omega}]_{RS}}{\delta \hat{\omega}^{AB}}
        \end{gathered}
    \end{equation}
    Then, we can fully expand the exterior derivative in differentials to arrive at the following set of conditions:\footnote{ Here, $(P_{(\beta)})^{IJ}_{KL}$ denotes the component expression of the map $\star+\beta$ on the Lie algebra. }
    \begin{equation}
    (P_{(\beta)})^{IJ}_{KL}B_{IRD}\theta_J = 0 \qquad \forall[RD]\neq [KL]
    \end{equation}
    \begin{equation}
    (P_{(\beta)})^{IJ,RS}U_{RSK}\theta_J = 0 \qquad \forall I \neq K
    \end{equation}
    \begin{equation}
        \left[
        A_{IK}(P_{(\beta)})^{IJ}_{AB} - \eta_{IK}(P_{(\beta)})^{IJ,RS}V_{RSAB}
        \right]\theta_J = 0 \qquad\forall K,[AB]
    \end{equation}
    We note that $(P_{(\beta)})^{IJ,RS}= (\star+\beta)^{IJ,RS}$ can be inverted to simplify these conditions. The third can be solved this way for $A_{IK}$ as a function of $V_{RSAB}$. Therefore, there are stringent constraints on which transformations can be interpreted as canonical ones. However, as can be verified by a straightforward but tedious calculation, for example the Lorentz transformations
    \begin{equation}
        X_\alpha = -\alpha\cdot\theta \frac{\delta}{\delta\theta} + d_{\hat{\omega}}\alpha \frac{\delta}{\delta\hat{\omega}}
    \end{equation}
    are symplectic, showing that some relevant vector fields can indeed satisfy these conditions.\\
    The vector fields \ref{STVectorFields} in the main text are not symplectic in general by this measure, as $B_{IRD}\neq 0$. An exception may exist though, as $\mathbb{T}_\phi$ may contain pieces that cancel the relevant dependence on $\hat{\omega}$. In that situation, the tetrad's transformation would be
    \begin{equation}
        X_\phi[\theta] = d_{\gamma[\theta]}\phi
    \end{equation}
    which clearly does not depend on the connection. As it turns out, we can write
    \begin{equation}
        \mathbb{T}_\phi = -\frac{1}{3} \kappa\cdot \phi
    \end{equation}
    with the contorsion of a general connection, so indeed by choosing $s=3$, we have a vector field with precisely this behaviour:
    \begin{equation}
        X_{\phi,3}[\theta] = d_{\gamma[\theta]}\phi 
        \qquad 
        (X_{\phi,3}[\omega])_{(\beta)}\wedge\theta = 2 (F_\omega)_{(\beta)}\cdot\phi
    \end{equation}
    This, at first, appears to be the only symmetry vector field among them that has a chance at being symplectic. In general, though, they all feature a \textit{bulk} nonintegrability. This is similar to the case of timelike diffeomorphisms, as the bulk obstructions vanish on-shell (on-shell of the Gauss constraint for $s=0$). Unlike diffeomorphisms, however, there is no clear interpretation of these terms.

\section{Extended phase spaces for integrability}
    The lack of immediate integrability could make some of our readers squeamish. In general, for a charge to be useful, one would like a generator which is general enough. From this point of view, it is well-motivated to seek extensions of the phase space that allow the above charges to be integrable. To make this clear, let us first illustrate what we mean in the example of the phase space of BF theory, equipped with the symplectic form
    \begin{equation}
        \Omega_\Sigma = \int_\Sigma \delta B\wedge\delta\omega.
    \end{equation}
    As stated above, the Kalb-Ramond translation charges \ref{BFKRC} for it have similar form to our new charges, so the analogy will be useful. In particular, they enjoy a centrally extended commutative Poisson algebra
    \begin{equation}
        \{K_\mu,K_{\Tilde{\mu}}\} = - \Lambda\oint_{\partial\Sigma} \Tilde{\mu} \wedge \Psi(\mu)
    \end{equation}
    In terms of on-shell charges, this suggests that the connection $\omega$ becomes noncommuting on the corner $\partial\Sigma$ when $\Lambda\neq 0$. This \say{change in commutation relation} is a generic phenomenon for gauge theories and may be puzzling at first, but can be clarified easily. Let us extend the phase space by a \textit{corner connection} $\eta$ supported only on $\partial\Sigma$, with appropriately extended symplectic form
    \begin{equation}
        \Omega_\Sigma = \int_\Sigma \delta B\wedge\delta\omega - \frac{1}{2\Lambda} \oint_{\partial\Sigma} \delta\eta\wedge\Psi^{-1}(\delta\eta).
    \end{equation}
    We can also extend manually the Kalb-Ramond charge to this phase space as
    \begin{equation}
        K_\mu = -\int_\Sigma \mu\wedge (F_\omega-\Lambda \Psi(B)) - \oint_{\partial\Sigma} \mu \wedge (\omega-\eta)
    \end{equation}
    and require that this vanishes for \textit{all} parameters on-shell. In this way, the connections become identified on-shell on the corner. Additionally, it makes the connection piece in the corner term into a difference of connections, which is a Lorentz tensor. Through adding this contribution from $\eta$, the full charge becomes Lorentz covariant. Moreover, though, it also becomes commutative:
    \begin{equation}
        \{K_\mu,K_{\Tilde{\mu}}\} = 0 
    \end{equation}
    The role of the original corner charges is now played by the pure corner charges
    \begin{equation}
        Q_\mu = \oint_{\partial\Sigma} \mu\wedge\eta \quad \{Q_\mu, Q_{\Tilde{\mu}} \} = \Lambda\oint_{\partial\Sigma} \Tilde{\mu} \wedge \Psi(\mu).
    \end{equation}
    We have thus separated out the noncommuting part of the connection cleanly at the kinematical level by introducing auxiliary corner data. Here, though, this procedure was optional, while for tetrad gravity we might even need it for integrability. The optionality in this case is encoded in the fact that there are no extra degrees of freedom on-shell due to the identification $\omega = \eta$.\\
    The first question for tetrad gravity is which variables to add. The formal similarity to the BF charges suggests having at least a corner connection, which would allow for a corner piece of the form
    \begin{equation}
        (\phi\lor\theta)_{(\beta)}\wedge (\omega-\eta)
    \end{equation}
    which is again Lorentz covariant. In other works\cite{freidelLoopGravityString2017}, a corner \textit{tetrad} $Z^I$ (really a zweibein) was introduced
    as a means to impose a boundary continuity equation like
    \begin{equation}
        \theta^2_{(\beta)} = Z^2_{(\beta)}.
    \end{equation}
    This could appear, for example, in charges of the form
    \begin{equation}
        C_\alpha = \int_\Sigma \theta^2_{(\beta)}\wedge d_\omega\alpha - \oint_{\partial\Sigma} Z^2_{(\beta)}\cdot \alpha
    \end{equation}
    A final possibility comes from including a \say{radial internal vector} $V^I$ into the corner phase space which, together with a corner zweibein, would allow the reconstruction of the bulk slice tetrad $\theta$ in the radial directions. For simplicity, we do not consider this here.\\
    Even if we include all of these into our corner phase space, we have not yet specified anything about their commutation relations. We could make a conservative Ansatz like
    \begin{equation}
        \Omega_{\partial\Sigma} =\oint_{\partial\Sigma}\omega_2 = \oint_{\partial\Sigma} A_I \delta\theta^I + B_I \delta Z^I + C\wedge \delta\eta 
    \end{equation}
    with vector-valued 1-forms $A,B$, vector valued 2-form $D$ and a Lie algebra valued 1-form $C$, which are all 1-forms on phase space. In this, we do not yet assume anything about them, so the form is not a priori nondegenerate. \\
    At first, then, we only require that the extended charge for translations is differentiable, meaning that
    \begin{equation}
        -\phi\cdot\delta\alpha + \delta(I_{Y_\phi}\omega_2) = d b^\phi_3
    \end{equation}
    for some completely arbitrary phase space function $b^\phi_3$. Then, really all we need to focus on are the $\delta\omega$-$\delta\theta$ terms to cancel the part coming from the bulk.\\
    We can further require of our extended charges to fulfil the same commutation relations as the pure bulk ones. This fixes, in particular, the transformation behaviour of the new variables $Z$ and $\eta$ under the bulk gauge transformations.
    In particular, requiring Lorentz covariance implies the simple law
    \begin{equation}
        X_\alpha[Z] = -\alpha\cdot Z \quad X_\alpha[\eta] = d_\eta\alpha
    \end{equation}
    so that $Z$ is a vector and $\eta$ a connection. For the translation behaviour, things are more complicated and we instead only have the highly implicit relation
    \begin{equation}
        (\Tilde{\phi}\lor \theta)_{(\beta)}\cdot Y_\phi[\eta] = - (Y_{\Tilde{\phi}}[\theta]\lor\phi)_{(\beta)}\cdot(\omega-\eta)
    \end{equation}
    and
    \begin{equation}
        Y_\phi[Z\wedge Z] = d_\omega(\phi\lor\theta) - \frac{1+\beta\star}{1+\beta^2}(\phi\lor \omega_{(\beta)}\wedge\theta)
    \end{equation}
    which shows that the corner frame needs to transform \textit{almost} in the same way as the bulk one. In principle, then, by assuming
    \begin{enumerate}
        \item Given form of the charges $P_\phi,C_\alpha$
        \item Differentiability (implying one can use the antisymmetric flow bracket for the charges)
        \item Covariance of the charges under the bulk gauge transformations
    \end{enumerate}
    we can constrain the possible form of a corner symplectic form for the new variables, as it must satisfy (similarly for Lorentz)
    \begin{equation}
        I_{Y_\phi}\Omega + \delta P_\phi = 0
    \end{equation}
   of which all elements except the corner symplectic form are fixed. However, the constraints are still too weak to uniquely identify the forms $A,B,C$. In particular, terms of the form
   \begin{align}
       A_I &= \dots +A_{I,J} \delta\theta^J \\
       B_I &=  \dots + B_{I,J} \delta Z^J \\
       C_{IJ} &=  \dots + C_{IJ,KL} \delta\eta^{KL}
   \end{align}
    may be added without influencing the above conditions. If these contributions are excluded, then we may in principle solve for the symplectic form\footnote{The coefficients of the symplectic form enter the conditions above in the general shape of a matrix equation $M\cdot x = b$ where $x$ contains the unknown coefficient functions, $M$ contains the coefficients of the vector fields $X,Y$ and $b$ contains the leftover bulk terms that need to be compensated. Generically, then, we expect that over most points in phase space $M$ is invertible, and therefore the symplectic form is uniquely determined.}, but its explicit form is most likely unilluminating.\\
    In principle, therefore, we can add a corner symplectic form to make the \textit{extended} charges, which include continuity laws, differentiable (on the part of phase space where one can solve for the symplectic form).
    We note, however, that this procedure is not as useful as it may first appear. Similarly to the idea of extended phase spaces, one has tremendous freedom in modifying the phase space of a theory artificially, as we did here. This freedom may be constrained by a choice of variables and further conditions, but as above, will leave open further freedom. Additionally, if such a modification is made, it must be justified for a certain end. Our original hope in this section was to somehow make the vector field $Y_\phi$ integrable. However, on an arbitrary extension of the phase space, this is always possible, so the question is meaningless by itself. In fact, if the extension presented here were to turn out to be totally unrelated to gravity, we would dismiss it immediately. Instead, a more \say{meaningful} question could be to \textit{not} extend the physical phase space and make only minor alterations in order to achieve integrability. However, we expect that any such alteration would simply amount to imposing boundary conditions implying \ref{IntegrabilityCondition}. 
    
    \section{Alternative approaches}
    We have so far highlighted that even from a purely spacetime covariant viewpoint\footnote{By this, we mean the approach of looking at the spacetime covariant transformations coming from ECH theory directly.}, there are potentially multiple choices of a \say{symmetry transformation} that mimicks the 3D Kalb-Ramond translations in various aspects. However, there are potentially even more options for such generalisations once we relax the requirement of working with Noether identities or spacetime diffeomorphisms directly. In this section, we will shed some light on three other ways to generalise the structures of the 3D case and give arguments for why these extensions are technically equally as cumbersome while providing fewer benefits for studying gravity.
    
    \subsection*{Covariant Ashtekar translations}\label{App:Ashtekar}
    The first alternative strategy one might consider is a direct continuation of translations found on the phase space of GR parametrised by Ashtekar's variables. Instead of recapping the extant literature, we will directly proceed to the implementation and highlight notable differences.\\
    To begin, we must make use of the \textit{adapted normal} $u_\Sigma$ we previously alluded to.
    We perform decompositions with respect to the field-dependent adapted normal $u_\Sigma$ to the spacelike slice $\Sigma$ as follows.
    Any internal vector $V^I$ has decompostion
    \begin{equation}
        V^I = V_\parallel^I + u_\Sigma^I V_\perp \qquad V_\perp = V\cdot u_\Sigma
    \end{equation}
    and Lorentz tensors $M^{IJ}$ split into two vectors
    \begin{equation}
    \begin{gathered}
                M = (M_\perp\lor u_\Sigma) - \star(M_\parallel\lor u_\Sigma)\\
                M^I_\perp = M^{IJ}u_{\Sigma} \qquad M^I_\parallel = (\star M)^{IJ}(u_\Sigma)_J
    \end{gathered}
    \end{equation}
    Meanwhile, if, for the given (spacelike) slice $\Sigma$, an associated normal 1-form $\mathbf{n}$ is available, together with some dual vector $\hat{n}$ satisfying $\mathbf{n}(\hat{n})=1$, any differential form $B$ can be decomposed into a horizontal piece (denoted by a tilde) and a vertical piece (denotes by an index n for normal to the slice):
    \begin{equation}
        B = \Tilde{B} + \mathbf{n}\wedge B_n\qquad B_n = i_{\hat{n}}B
    \end{equation}
    Similarly, for (multi)vector fields $\xi$,
    \begin{equation}
        \xi = \xi_\perp + \hat{n}\wedge \xi_n \qquad \mathbf{n}(\xi)=\xi_n.
    \end{equation}
    Now let us do this for tetrad gravity. The tetrad decomposes as
    \begin{equation}
		\theta = \Tilde{e}+\mathbf{n} u_\Sigma \qquad \Tilde{e}\cdot u_\Sigma = 0,
    \end{equation}
    the spin connection as
    \begin{equation}
        \begin{gathered}
            \omega 
            = ( K\lor u_\Sigma) + \Gamma \qquad d_\Gamma u_\Sigma = 0, K=d_\omega u_\Sigma \\
            = ( (K-d u) \lor u ) - \star (\Gamma_\parallel\lor u)
        \end{gathered}
    \end{equation}
    and the curvature as
    \begin{equation}
        F_\omega = (d_\Gamma K\lor u) + F_\Gamma -(K \wedge K).
    \end{equation}
    We can then give a connection $\acal$ satisfying the following properties:
	\begin{enumerate}
		\item It is a Lorentz connection for the original Lorentz group.
		\item It is Poisson commuting.
		\item Its Poisson brackets with triads are such that its holonomies can be used to construct a holonomy-flux algebra.
		\item It transforms covariantly under spatial diffeomorphisms, so those with $i_{\xi} \mathbf{n}=0$.
		\item It reduces, when $u_\Sigma$ takes the value $\delta^I_0$, to the Ashtekar-Barbero-Sen (ABS) connection.
		\item It makes the internal normal $u_\Sigma$ covariantly constant, $d_\acal u_\Sigma = 0$. This means that it has no extrinsic curvature, and in particular that its curvature has no perpendicular components.
		\item It satisfies $\omega_\perp-\beta\omega_\parallel = \acal_\perp-\beta\acal_\parallel$.
	\end{enumerate}
	Then in fact, given just requirements 1, 2, 3 and 4 (particularly 2) (or alternatively, 1,6 and 7), there turns out to be a unique solution to this problem, also satisfying the other two properties.\cite{Alexandrov:2001wt,Alexandrov:2011ab,geillerLorentzCovariantConnectionCanonical2011a,dupuisLiftingSUSpin2010} \footnote{Dropping the commutativity opens up the possibility of spacetime covariance. In fact, this connection is the \textit{only} commutative one satisfying 2,3 and 4, and is not timelike diffeomorphism covariant. There is exactly one spacetime diffeomorphism covariant connection on this phase space, which on-shell of the Gauss constraint coincides with the spin connection.} It is given by
	\begin{align}
		\mathcal{A} 
		= \Gamma + \star(\gamma K\lor u_\Sigma)
	\end{align}
	We then have
	\begin{equation}
		\acal_\parallel = A_{ABS}= \Gamma_\parallel -\gamma K , \qquad \acal_\perp = -du_\Sigma
	\end{equation}
	So we have done the replacement 
	\begin{align}
		-du_\Sigma + K &\mapsto -du_\Sigma \\
		\Gamma_\parallel &\mapsto   \Gamma_\parallel -\gamma K
	\end{align}
	which means we removed the extrinsic curvature and reshuffled it into the rotation part of the connection. We can easily see that this is a Lorentz connection as $\Gamma$ is one and $K\lor u_\Sigma$ is a Lorentz tensor.\\
	The curvature of this connection is not directly related to the spin connection curvature we usually associate with gravity.
	\begin{equation}
		F_\acal = F_\Gamma +  \gamma\star (d_\Gamma K \lor u_\Sigma) + \frac{\gamma^2}{2} K\wedge K.
	\end{equation}
	Compare this to the spin connection, where
	\begin{equation}
		F_\omega = F_\Gamma +  (d_\Gamma K \lor u_\Sigma) -\frac{1}{2} K\wedge K.
	\end{equation}
	So flatness of $\acal$ does not imply that the spacetime metric associated to $\omega$ is flat.\\ 
    The spin connection itself may be reconstructed from the covariant ABS connection, as well. To begin, define the torsion of the ABS connection, also referred to as the \textit{ABS momentum}:
	\begin{equation}
		P^I = d_\acal \theta^I  
	\end{equation}
	which has the simple decomposition
	\begin{equation}
		P^I = d_A e^I +  u_\Sigma^I d\mathbf{n}
	\end{equation}
	This relates back to the original spin connection torsion as
	\begin{align}
		P^I &= T^I - (1-\gamma\star)(K\lor u_\Sigma)^{IJ}\wedge\theta_J\\
		&= T^I 
		+  \mathbf{n}\wedge K^I 
		+  u_\Sigma^I K^J\wedge e_J 
		-  \gamma (K\times e) 
	\end{align}
        where we define the cross product as $ A\times B := \star(A\lor B)\cdot u_\Sigma $.
	At least on-shell of Gauss, then, 
	\begin{equation}
		i_{\hat{n}}P^I_\parallel \approx K^I
	\end{equation}
	and we can reconstruct the spin connection as
	\begin{equation}
		\omega \approx \acal + (1-\gamma\star)(i_{\hat{n}}P\lor u_\Sigma)
	\end{equation}
	The other parts are for example
	\begin{equation}
		\Tilde{P}^I_\parallel = \gamma p^I = -  \gamma (K\times e) 
	\end{equation}
	which we call the \textit{spatial momentum aspect}\cite{freidelEdgeModesGravity2020a}, and
	\begin{equation}
		i_{\hat{n}}P_\perp =  K_n^J\wedge e_J \approx - \lcal_{\hat{n}}\mathbf{n} \qquad \Tilde{P}_\perp = \Tilde{K}^J\wedge e_J \approx 0
	\end{equation}
	so onshell of Gauss the first is the \textit{acceleration} of the slice, while the other vanishes.
    The ABS momentum will take on the role of the translation generator density. It is therefore already possible to see what happens for timelike translations $\phi\sim u_\Sigma$: They will have vanishing generator as $\Tilde{P}_\perp = 0$. This points to a crucial deficiency of the approach: It does not feature more than spatial translations.\\

    With this in mind, we can study the 3-dimensional translations of the frame
	\begin{align}
		T_\phi[\tilde{e}] &= d_\acal \phi \\
		T_\phi[\acal^{IJ}] &= - i_{\hat{e}_I}F_\acal^{JK}\phi_K \approx 0\\
		T_\phi[u_\Sigma] &= 0
	\end{align}
	for parameters $\phi$ with $\phi\cdot u_\Sigma = 0$, so with a field dependence reflected in the decomposition
	\begin{equation}
		\delta\phi = (\delta\phi)_\perp -   u_\Sigma (\phi\cdot\delta u_\Sigma).
	\end{equation}
	If we assume $(\delta\phi)_\perp = 0$,\footnote{or more generally for $P_I (\delta\phi)_\perp^I - d(\tilde{e}_I (\delta\phi)_\perp^I) = \delta f $} then these translations are integrable with generator
	\begin{equation}
		T_\phi = \int_\Sigma \beta P_I \wedge d_\acal\phi^I 
	\end{equation}
	which is onshell just
	\begin{equation}
		T_\phi \approx  \oint_{\partial \Sigma}\beta \tilde{e}_I  d_\acal\phi^I.
	\end{equation}
    For this, it is necessary and noteworthy that up to Boost constraint terms, the spatial diffeo constraint can be rewritten as
    \begin{equation}
        d_\acal P \approx 0.
    \end{equation}
    Therefore, the on-shell phase space of tetrad gravity can be equivalently achieved by imposing the full Gauss constraint, this spatial momentum constraint as well as the usual Hamiltonian constraint.
    Their Poisson brackets show a central term, already known from the literature, but also a corner term involving the curvature,
	\begin{equation}
		\{T_\phi,T_{\Tilde{\phi}}\} = -\oint_{\partial S_P}\beta(\phi\lor\Tilde{\phi})\cdot F_\acal + \oint_{\partial S_P}\beta\phi_I d\Tilde{\phi}^I.
	\end{equation}
	 The translations are therefore bulk-commuting.\\
    We can once again see that the \textbf{timelike translation charges vanish identically}. This is the main drawback of this approach - it does not allow for uniform/Lorentz covariant treatment of the charges and transformations. However, we see the appearance of corner noncommuting curvature-dependent terms as a possibly generic feature.

    \subsection*{Reverse engineered translations}
        Here, we will approach the issue of translation charges and particularly their covariant form from the opposite side: If we \textit{assume} the sought-after type of translations to form a vector representation of the Lorentz group, we can actually find translation charges directly. The idea is simple: Knowing an expression for the Lorentz charges on the tetrad gravity phase space,
    \begin{equation}
        \int_\Sigma \theta^2_{(\beta)} \wedge d_\omega \alpha,
    \end{equation}
    we can apply a frame translation to it. What then goes into this strategy as an ingredient is the specific form of transformation, e.g. $\theta\mapsto\theta +d_\omega \phi$. If these translations are faithfully represented on the phase space by Hamiltonian vector fields, then we should have that the change is given by
    \begin{equation}
        \{T_\phi,C_\alpha\} = T_{\alpha\cdot\phi}
    \end{equation}
    It must be this if indeed the charges are vectors under Lorentz transformations. Then, if we can write an arbitrary parameter $\Phi$ through some $\phi,\alpha$, we can reconstruct the appropriate charges $T_\Phi$ immediately. Such parameters are easy to achieve for the Poincare algebra; just choose
    \begin{equation}
        \alpha = \frac{\Phi \lor\phi}{\phi^2}
    \end{equation}
    with some (normalised) reference vector $\phi$ which is orthogonal to $\Phi$.\\

    What this approach finds is a hint towards the form of these charges. The expressions found look generically like corner charges from the other strategies, but particularly from the covariant translations. If one tries to specifically find the expression for timelike translation corner charges, one can then find that they must all vanish on-shell. In comparison, in the covariant translation strategy, the charge is
    \begin{equation}
        \int_{\partial\Sigma} \phi_\perp (\Gamma_\parallel + \beta K)_I \Tilde{e}^I 
    \end{equation}
    of which the $K-e$ term certainly matches and vanishes. The additional $\Gamma-e$ piece, though, is nonzero and perhaps the true charge we are looking for. Overall, this approach, while almost algorithmic, has the two immediate deficits of requiring to know the transformation law a priori, and also that it only produces the corner charge. What it however establishes is that, if only the naive tetrad translations are used, one cannot find nontrivial time translation corner charges.

\bibliography{P3S1}

\end{document}